\documentclass[12pt]{article}

\usepackage{pgfplots}
\usetikzlibrary{decorations.pathmorphing}

\tikzset{snake it/.style={decorate, decoration=snake}}
\pgfplotsset{compat=1.10}
\usepgfplotslibrary{fillbetween}
\usetikzlibrary{patterns}

\usepackage{draft} 
\usepackage{hyperref}
\usepackage{graphicx,color,subfig}
\usepackage{cite}
\usepackage{mciteplus}
\usepackage{skak}
\usepackage{xcolor}
\usepackage{empheq}
\usepackage{tikz}
\DeclareFontFamily{OT1}{pzc}{}
\DeclareFontShape{OT1}{pzc}{m}{it}{<-> s * [1.10] pzcmi7t}{}
\DeclareMathAlphabet{\mathpzc}{OT1}{pzc}{m}{it}

\def\be#1\ee{\begin{align}#1\end{align}}



\newcommand{\N}{{\cal N}}
\newcommand{\M}{{\cal M}}

\newcommand{\beq}{\begin{eqnarray}}
\newcommand{\eeq}{\end{eqnarray}}
\begin{document}

\unitlength = .8mm

\begin{titlepage}

\begin{center}

\hfill \\
\hfill \\
\vskip 1cm

\title{\Huge $\N=2$ Heterotic Strings Revisited}

\author{Amit Giveon$^1$, Akikazu Hashimoto$^2$, David Kutasov$^3$}
\address{
$^1$Racah Institute of Physics, The Hebrew University \\
Jerusalem, 91904, Israel\\
$^2$Department of Physics, University of Wisconsin\\
Madison, WI 53706, USA\\
$^3$Kadanoff Center for Theoretical Physics and Enrico Fermi Institute\\ 
University of Chicago, Chicago IL 60637
}
\vskip 1cm

\email{$^1$giveon@mail.huji.ac.il, $^2$aki@physics.wisc.edu, $^3$dkutasov@uchicago.edu}

\end{center}

\abstract{We show that $1+1$ dimensional vacua of the $(0,2)$ heterotic string have many properties in common with more recently studied physical  systems. The free theory exhibits a symmetric product CFT structure. The single-string Hilbert space splits into sectors labeled by the string winding, $w$, that contain purely left-moving (for $w>0$) and right-moving (for $w<0$) excitations. These sectors exhibit infinite-dimensional left and right-moving symmetry algebras, respectively, with central extensions that depend on $w$. String interactions between  left and right-movers appear to lead to a $T\bar T$ deformation of the free theory. 
}

\vfill

\end{titlepage}

\eject

\begingroup
\hypersetup{linkcolor=black}
\tableofcontents
\endgroup

\vskip 2cm

\section{Introduction}

In this paper, we revisit $\N=2$ heterotic strings, focusing on their $1+1$ dimensional vacua. These theories were first studied in the 1990's \cite{Ooguri:1991fp,Ooguri:1991ie,Kutasov:1996fp,Kutasov:1996zm,Kutasov:1996vh}, and were shown to exhibit interesting properties. As we will see, some of their properties are related to more recently studied topics, such as symmetric product CFT, and $T\bar T$ deformed CFT. They also share many properties with string theory on $AdS_3$ (with NS $B$-field). For example, we will see that they have the property that string winding around the spatial direction plays a central role in their analysis. The winding zero sector contains generators of an infinite-dimensional spacetime symmetry algebra, while the $w\not=0$ sectors contain states that transform non-trivially under that symmetry. The symmetry includes left and right-moving Kac-Moody and Virasoro algebras, whose action on sectors with $w\not=0$ has central extensions that depend on $w$. Another property that $\N=2$ heterotic strings share with string theory on $AdS_3$ is that their spacetime properties mirror in many ways the worldsheet ones. 

The relations described in the previous paragraph are one of the main motivations for this study. We hope that further studying $\N=2$ strings will shed light on the systems mentioned above, and vice versa. Another motivation is the work of \cite{Kutasov:1996fp,Kutasov:1996zm,Kutasov:1996vh}, which suggested that $\N=2$ heterotic strings are relevant for a deeper understanding of string dualities. That work involved $(1,2)$ heterotic strings, but in this paper we will focus on the $(0,2)$ theory, postponing the analysis of the $(1,2)$ one to another publication. 

It was recognized in the 1990's that $\N=2$ heterotic strings can be formulated in $1+1$ or $2+1$ dimensions. In this paper, we  focus on the $1+1$ dimensional construction, leaving the $2+1$ dimensional one to future work. A major theme in our discussion will be the difference between the case where the spatial direction is non-compact, and the one where it is compactified on a circle. We will start with the former, and then turn to the latter.

The plan of the remainder of the paper is the following. In section \ref{noncomp}, we describe the basic structure of the $(0,2)$ heterotic string and some of its properties, in the case where the $1+1$ dimensional target space is non-compact $(\mathbb{R}^{1,1})$. This serves to fix our notation and define various quantities that enter the analysis. In particular, we define the Niemeier CFT $\M$ -- a holomorphic $c=24$ worldsheet CFT that enters the construction of a class of vacua of the $(0,2)$ heterotic string. We show that the spectrum of the theory is given by a set of scalar fields, $\phi^a$, that transform in the adjoint representation of the Lie algebra $G$ associated with $\M$. 
The spacetime theory for the fields $\phi^a$, restricted to the Cartan subalgebra (CSA) of $G$, appears to be a $T\bar T$ deformation of a free field theory, with $T\bar T$ coupling $\sim g_s^2\alpha'$. 

In section \ref{comp}, we compactify the spatial direction in target space on a circle of radius $R$. We focus on a new feature of this case -- the appearance of sectors in the Hilbert space that describe strings winding multiple times around the circle. We show that for positive winding, the single-string spectrum contains only left-moving excitations on the string, while for negative winding, all the excitations are right-moving. For $w=1$, the spectrum coincides with that of the Niemeier CFT, $\M$, that went into its worldsheet construction. Similarly, for $w=-1$, we find a copy of the right-moving analog of $\M$. For $|w|>1$, we find a single-string spectrum that, as we show in section \ref{symmprodn}, agrees with that of the $\mathbb{Z}_{|w|}$ twisted sector in the corresponding symmetric product of $\M$. 

We also show in section \ref{comp} that the spacetime theory has an interesting algebraic structure. The $w=0$ sector contains left and right-moving affine Lie algebras, $\widehat G_L$, $\widehat G_R$, associated with the Niemeier CFT $\M$. $\widehat G_L$ acts naturally on states with $w>0$, while $\widehat G_R$ acts on states with $w<0$. The central extensions of both algebras depend on the winding sector. These algebras act as spectrum generating algebras on the single-string spectrum, arranging it into representations.

We also find in section \ref{comp} that the $w=0$ sector contains left and right-moving higher spin symmetry generators, and in particular Virasoro algebras, with similar properties. The central extensions again depend on the winding sector, and the symmetry generators act on the $w\not=0$ sectors as spectrum generating operators. 

In section \ref{torus}, we study the thermal partition sum of the $(0,2)$ string theory. We show that the spectrum of section \ref{comp} can be recovered from this partition sum. In section \ref{symmprodn}, we study some properties of the symmetric product of Niemeier CFT's, $\M^N/S_N$. In particular, we describe its spectrum and symmetry structure, and calculate its thermal partition sum, in preparation for comparing them to the spacetime theory of the $(0,2)$ string in section \ref{spdyn}. In section \ref{discuss}, we comment on our results and possible extensions. Two appendices contain the derivation of some technical results used in the text.

\section{Non-compact theory} \label{noncomp}

In the $(0,2)$ heterotic string, the left-moving worldsheet degrees of freedom are those of the bosonic string, while the right-movers are those of the $\N=2$ string. Thus, the right-moving worldsheet CFT is described by four scalars $x^\mu$, $\mu=0,1,2,3$, with signature $(-,-,+,+)$, and their four superpartners under right-moving worldsheet $\N=2$ superconformal symmetry, the fermions $\bar\psi^\mu$. We take the scalars $x^\mu$ to be non-compact, though later we will compactify some of them. 

The right-moving $\N=2$ superconformal generators are given by 
\begin{equation}
	\label{ntwo}
 \begin{split}
    \bar T=&-\frac1{\alpha'}\bar\partial x^\mu\bar\partial x_\mu-\frac12\bar\psi^\mu\bar\psi_\mu~,\\
\bar G^\pm=&\frac{i}{\sqrt{2\alpha'}}(\eta^{\mu\nu}\pm {\cal I}^{\mu\nu})\bar\psi_\mu\bar\partial x_\nu~,\\
\bar J=&\frac12{\cal I}^{\mu\nu}\bar\psi_\mu\bar\psi_\nu~.
\end{split}
	\end{equation}
Here, the metric is $\eta^{\mu\nu}={\rm diag}(-1,-1,1,1)$, and the complex structure ${\cal I}^{\mu\nu}$ satisfies 
${\cal I}^{\mu\nu}=-{\cal I}^{\nu\mu}$, ${\cal I}_{\mu\nu}{\cal I}^{\nu\lambda}=\eta_\mu^\lambda$. We will take ${\cal I}^{03}={\cal I}^{12}=1$.

$\N=2$ worldsheet supersymmetry is gauged in the $\N=2$ string. In the superconformal gauge, one finds two $(\bar\beta,\bar\gamma)$ systems, that couple to $\bar G^\pm$, \eqref{ntwo}, $(\bar\beta_\pm,\bar\gamma_\pm)$. From them, we can obtain in the usual way two bosons $\bar\varphi_\pm$, $\bar\beta_\pm\bar\gamma_\pm=\bar\partial\bar\varphi_\pm$. 
These play the same role as in the ${\cal N}=1$ fermionic string, giving rise to the notion of pictures, as we review below. 

The left movers are those of the bosonic string. Since the scalars $x^\mu$, $\mu=0,1,2,3$, are non-compact, they have left-moving components as well, and contribute to the left-moving stress-tensor,
\begin{equation}
	\label{leftxmu}
T_x=-\frac1{\alpha'}\partial x^\mu\partial x_\mu \ .
	\end{equation}
Normally, that would mean that we need to add to \eqref{leftxmu} twenty two left-moving scalars, but in this case, it is more convenient to proceed in a different way \cite{Ooguri:1991ie}. We add to the $x^\mu$ twenty four left-moving scalars $y^j$, $j=1,2,\cdots, 24$, which contribute to the left-moving stress tensor\footnote{It is convenient to take $y^i$ to be canonically normalized, $y^i(z)y^j(0)\sim -\delta^{ij}\ln z$, whereas  $x^\mu$ \eqref{leftxmu} have the standard normalization \cite{Polchinski:1998rq}, $x^\mu(z) x^\nu(0)\sim-\frac{\alpha'}2\eta^{\mu\nu}\ln|z|^2$. Thus, the worldsheet left-moving scaling dimension of $\exp(i\vec p\cdot\vec y)$ is $\Delta_L=\frac12|\vec p|^2$, while for $\exp(ip_\mu x^\mu)$ we have $\Delta_L=\Delta_R=\frac{\alpha'}4p_\mu p^\mu$.} 
\begin{equation}
	\label{leftyi}
T_y=-\frac12\partial y^j\partial y^j \ .
	\end{equation}
Thus, the left-movers appear to live in  a $26+2$ dimensional spacetime, with $(x^0,x^1)$ the two time dimensions, and $(x^2, x^3, \vec y)$ the twenty six space dimensions. 

In order to reduce the left-moving spacetime to $25+1$ dimensions (the critical dimension and physical signature of the bosonic string), we gauge a left-moving null $U(1)$ current. There are two qualitatively different ways of doing that \cite{Ooguri:1991ie}. If we choose this current to lie in the $\mathbb{R}^{2,2}$ parametrized by $x^\mu$, we get a theory that lives in $1+1$ non-compact dimensions. On the other hand, if the null $U(1)$ involves one of the compact directions $\vec y$, the resulting theory is $2+1$ dimensional. 

As mentioned above, in this paper we will study the $1+1$ dimensional vacua of the $(0,2)$ heterotic string. We will take the null $U(1)$ current to be $J=\partial x_0+\partial x_3$ (w.l.g). The null gauging effectively removes the two worldsheet fields $(x^0, x^3)$, so the resulting theory lives in $1+1$ dimensions, 
\begin{equation}
	\label{deftx}
	(x^1, x^2)\equiv (t,x)\ .
	\end{equation}
The twenty four left-moving scalars $\vec y$ must be described by a holomorphic modular invariant worldsheet CFT. This means that the momentum $\vec p_L$ for these scalars lives in an even, self-dual lattice, one of the twenty four Niemeier lattices. We will refer to this holomorphic theory as a {\it Niemeier CFT}, and will denote it by $\M$.  

We now turn to the spectrum of the resulting string theory. Since the right-movers are those of the $\N=2$ string, we can proceed as in the $\N=1$ superstring. In the $(-1, -1)$ picture for the two ghost systems mentioned above, the right-moving part of the physical vertex operators takes the form 
\begin{equation}
	\label{rightm}
	e^{-\bar\varphi_+-\bar\varphi_-}e^{ip_\alpha  x^\alpha}=e^{-\bar\varphi_+-\bar\varphi_-}e^{-iEt+ipx} \ .
	\end{equation}
Here, $\alpha=1,2$ runs over the two spacetime directions\footnote{We can set $p_0=p_3=0$ using the null gauging.} $(t,x)$, \eqref{deftx}, and $(E,p)$ are the components of $p^\alpha$. They are continuous in the non-compact theory, and satisfy the mass-shell condition  
\begin{equation}
	\label{masss}
	p_\alpha p^\alpha=0,\;\; {\rm i.e.}\;\; E=\pm p \ .
	\end{equation}
On the left-moving side, to satisfy the mass-shell condition \eqref{masss}, we need to add to \eqref{rightm} a dimension $(1,0)$ operator, which we will denote by $J^a$, from the Niemeier CFT $\M$. The $J^a$ include the currents $i\partial y^j/\sqrt2$, $j=1,\cdots, 24$, and operators of the form $\exp(i\vec p_L\cdot \vec y)$, with $\vec p_L$ any vector of length squared $|\vec p_L|^2=2$ in the Niemeier lattice. Together, they give the familiar vertex operator construction of a rank twenty four simply laced affine Lie algebra $\widehat G$, with $i\partial y^j/\sqrt2$ giving the CSA generators, and $\vec p_L$ being the roots of $G$ \cite{Goddard:1986bp}. They satisfy the OPE algebra 
\begin{equation}
	\label{wsKM}
	J^a(z)J^b(\xi) \sim \frac{ k\delta^{ab}/2}{(z-\xi)^2}+\frac {i f^{ab}_{\,\,\,\,\,\,c} J^c}{z-\xi}~,
	\end{equation}
where $f^{abc}$ are the structure constants of the Lie algebra $G$. The level $k$ of this algebra is equal to one \cite{Goddard:1986bp}. Thus, while we include the level $k$ in many subsequent equations, it should always be understood to take the value $k=1$ below. 

 The full vertex operator in the $(0,2)$ string, in the $(-1,-1)$ picture for the right-movers, takes the form  \begin{equation}
	\label{fullv}
	V^a(k)=\int d^2z J^a e^{-\bar\varphi_+-\bar\varphi_-}e^{ip\cdot x} \ .
	\end{equation}
The mass-shell condition \eqref{masss} implies that \eqref{fullv} describes $\dim(G)$ massless scalar fields, $\phi^a(x_\alpha)$. 

We will refer to operators with $E=p$ as left-moving, and those with $E=-p$ as right-moving. In terms of lightlike coordinates and momenta, 
\begin{equation}
	\label{xpm}
	x^\pm=t\pm x, \;\;\; p^\pm=\frac12(E\pm p)~,
	\end{equation}
the left-moving vertex operators are given by  
\begin{equation}
	\label{fullvL}
	V_L^a(p^+)=\int d^2z J^a e^{-\bar\varphi_+-\bar\varphi_-}e^{ip^+ x^-} \ .
	\end{equation}
$V_R$ is given by a similar expression with $+\leftrightarrow -$. 

As usual, for calculations we also need the $(0,0)$ picture version of the operators \eqref{fullv}, \eqref{fullvL}. To find it, we need to act on \eqref{fullv},\eqref{fullvL} with $\bar G_{-\frac12}^+$ and $\bar G_{-\frac12}^-$ \eqref{ntwo}. One of the two always gives zero when acting on the operator \eqref{fullv}. Thus, the action of $\bar G_{-\frac12}^+$ and $\bar G_{-\frac12}^-$ on the $(-1,-1)$ picture operator \eqref{fullv} is the same as the action of their anti-commutator, $\{\bar G_{-\frac12}^+,\bar G_{-\frac12}^-\}=\bar L_{-1}$.

We conclude that the $(0,0)$ picture version of the vertex operator \eqref{fullv} is 
\begin{equation}
	\label{fullv0}
	V^a(p)=\int d^2z J^a \bar\partial e^{ip\cdot x}\ .
	\end{equation}
This is a familiar vertex operator from studies of the standard, $(0,1)$, heterotic strings. It describes a $1+1$ dimensional gauge field with polarization proportional to the momentum, in other words, a pure gauge one. The scalar fields $\phi^a$ mentioned after eq. \eqref{fullv} can be thought of as the gauge functions, related to the gauge fields as 
\begin{equation}
	\label{aaa}
	A_\alpha^a(t,x)=\partial_\alpha \phi^a(t,x) \ .
	\end{equation}
The mass-shell condition \eqref{masss} can be thought of as the Lorenz gauge condition $\partial^\alpha A_\alpha^a=0$, which leads to the massless Klein-Gordon equation for $\phi^a$, $\partial_\alpha\partial^\alpha\phi^a=0$. 

One might expect that a pure gauge field such as \eqref{aaa} is trivial, and thus the physics should be independent of $\phi^a$. However, as is familiar from other contexts, this is the case if the gauge function goes to zero as we approach the boundary of spacetime. Gauge functions that do not go to zero at the boundary do not decouple in general. Gauge functions that go to a constant at the boundary are expected to give rise to global symmetries of the model, while ones that diverge as we approach the boundary, give rise to a change of the background. 

We will find below that these expectations are realized. In particular, we will argue that one should think of the physical states that appear in the above perturbative string analysis as modes of currents that form an infinite symmetry algebra, and that certain solutions for $\phi^a(x)$ that diverge near the boundary give rise to a Narain-type vacuum manifold. 

Consider, for example, the left-moving vertex operator \eqref{fullvL}. The corresponding zero-picture vertex operator \eqref{fullv0} takes the form 
\begin{equation}
	\label{zerovL}
	V_L^a(p^+)=\int d^2z J^a \bar\partial e^{ip^+ x^-} \ .
	\end{equation}
The operator \eqref{zerovL} is labeled by a momentum, $p^+$. It's instructive to Fourier transform it, and define a position space operator, 
\begin{equation}
	\label{possp}
	V_L^a(x)=\int d^2z J^a \bar\partial \delta(x^-(z,\bar z)-x) \ .
	\end{equation}
Note that in \eqref{possp}, $x^-(z,\bar z)$ is a worldsheet field, while $x$ is a constant, the Fourier transform of $p^+$. Also, in an abuse of notation, we denote the position-space operator \eqref{possp} by the same letter as the momentum-space one \eqref{zerovL}.  

The vertex operators \eqref{possp} are localized in $x^-$, but are independent of $x^+$. They correspond to gauge functions $\phi^a$ \eqref{aaa} that do not go to zero as $x^+\to\infty$. According to the discussion above, we expect them to give rise to symmetry generators. We next show that this is indeed the case. 

As we describe in more detail in section \ref{comp}, we can rewrite \eqref{possp} as 
\begin{equation}
	\label{curalg}
	V_L^a(x)=\oint \frac{dz}{2\pi i} J^a \delta(x^-(z,\bar z)-x) \ ,
	\end{equation}
where the contour integral runs over small circles around other insertions in a correlation function. In particular, we can calculate the commutator 
\begin{equation}
\label{commalg}
[V_L^a(x),V_L^b(y)]=\oint \frac{d\xi}{2\pi i}\oint_{\xi} \frac{dz}{2\pi i} J^a(z)\delta(x^-(z,\bar z)-x)J^b(\xi)\delta(x^-(\xi,\bar\xi)-y)\ .
\end{equation}
The $z$ integral in \eqref{commalg} is over a small contour around $\xi$. To calculate it, we use the worldsheet OPE algebra \eqref{wsKM}. We find
\begin{equation}
\label{curralgebra}
[V_L^a(x),V_L^b(y)]=
if^{ab}_{\,\,\,\,\,\,c}\delta(x-y)V_L^c(y)-
 \frac k2 \delta^{ab}\delta'(x-y)
 \oint \frac{d\xi}{2\pi i}\partial_\xi x^-\delta(x^-(\xi,\bar\xi)-y)
 ~\ .
\end{equation}
The commutation relation \eqref{curralgebra} is the famous current algebra with central extension. We will study it in more detail in section \ref{comp}.

The $(0,0)$ picture vertex operators \eqref{fullv0} can also be used in the sigma-model approach to string theory, to study the field space parametrized by the $\phi^a(x)$. To do that, one adds to the worldsheet action the term 
\begin{equation}
	\label{defac}
	\delta S=-i\int d^2z J^a \bar\partial \phi^a(x^\alpha) \ .
	\end{equation}
For general $\phi^a(x^\alpha)$, this perturbation breaks (worldsheet) conformal symmetry, however, setting 
\begin{equation}
	\label{modsp}
	\phi^j(x^\alpha)=\sqrt2C^j_\alpha x^\alpha~,
	\end{equation}
 where $j$ runs over the CSA of $G$ (and setting all the other $\phi^a$ to zero), 
gives rise to solutions for all values of the constants $C^j_\alpha$. Indeed, plugging \eqref{modsp} into \eqref{defac}, and recalling the definition of the currents $J^j$, we find that the deformation takes the form 
\begin{equation}
	\label{defac1}
	\delta S=C^j_\alpha\int d^2z \partial y^j \bar\partial x^\alpha~,
	\end{equation}
where the index $j=1,\cdots, 24$ runs over the twenty four dimensions of $\M$. The deformation \eqref{defac1} is well known to be truly marginal; it parametrizes a Narain-type moduli space labeled by the constants $C^j_\alpha$. 

Note that the above discussion is compatible with our statement that when the gauge functions $\phi^a$ \eqref{modsp} do not go to zero near the boundary of $\mathbb{R}^{1,1}$, they can have an effect on the theory. The gauge functions \eqref{modsp} in fact diverge at the boundary of $\mathbb{R}^{1,1}$, so it's not surprising that they have the effect of changing the background. Furthermore, the gauge functions \eqref{modsp} are not  good operators in the worldsheet theory,\footnote{They do not have powerlike correlation functions.} which is another reason that the deformation \eqref{defac} has an effect. 

The above discussion is also compatible with the observation in the literature from the 1990's, that the effective action of the $\phi^j$ is the Nambu-Goto action \cite{Kutasov:1996vh}, 
\begin{equation}
	\label{NGact}
	S\simeq -\frac{1}{\alpha'g_s^2}\int dx dt \sqrt{-\det g_{\alpha\beta}}~;\;\;\;
g_{\alpha\beta}=\eta_{\alpha\beta}+\alpha'\partial_\alpha \phi^j\partial_\beta\phi^j~,
	\end{equation}
where $g_s$ is the string coupling of the $\N=2$ string. 
The NG action \eqref{NGact} also has the solutions \eqref{modsp}, and in fact can be viewed as the effective action for fluctuations around the background with general $C^j_\alpha$. 

An interesting question is what are the observables in the $\N=(0,2)$ heterotic string. Usually, in string theory, the observables are S-matrix elements of physical vertex operators, which in our case are given by \eqref{fullv0}. In the $\N=2$ string these S-matrix elements are known to vanish. This is natural, since the vertex operators \eqref{fullv0} describe pure gauge fields, \eqref{aaa}. However, there may be other observables, which we will discuss in a future publication. Here we note that one observable is the action for the $\phi^j$ \eqref{NGact}. One way to think about this NG action is as determining the Zamolodchikov metric on the moduli space labeled by the $C^j_\alpha$ in \eqref{modsp}, \eqref{defac1}. 

It is instructive to compare the role of the NG action in our case to that of the DBI action for D-branes. In that case, the  moduli space analogous to \eqref{modsp} corresponds to rotations and boosts of the D-brane, as well as turning on a constant field strength for the worldvolume gauge field on the brane. The DBI action allows one to study small fluctuations on a D-brane, at a general point on this moduli space. From the worldsheet point of view, it is given by the disk partition sum in a state corresponding to a D-brane at a particular point in moduli space. It would be interesting to understand the analog of this in our case. Of course, D-branes have a richer spectrum of excitations, and one can study their dynamics using open string theory. This aspect of D-brane physics is more subtle in our case. 

We finish this section with a few comments:
\begin{itemize}
\item The action \eqref{NGact} is reminiscent of the one that appears in $T\bar T$ deformed CFT \cite{Smirnov:2016lqw,Cavaglia:2016oda}. Indeed, if we start with the free field theory of twenty four scalars $\phi^j$, its $T\bar T$ deformation with deformation parameter $g_s^2\alpha'$ gives the action \eqref{NGact} \cite{Cavaglia:2016oda}.
\item The coefficient multiplying the action \eqref{NGact} suggests that one can think of the spacetime theory as describing a large number, $N\sim 1/g_s^2$, of fundamental strings stretched in the $x$ direction. This is reminiscent of what happens in string theory in $AdS_3$, where the $AdS$ background is obtained by adding a large number of strings to a fivebrane throat. It also suggests that the spacetime theory should have a symmetric product structure. This will become clearer in the next sections.
\item Above, we discussed the action for the fields that belong to the CSA algebra of $G$, $\phi^j$. The analysis can be generalized to the non-abelian case, but we will not describe the details here.
\end{itemize}

\section{Compact theory} \label{comp}

In this section, we study the theory of section \ref{noncomp} in the case where the spatial coordinate $x$ \eqref{deftx} is compactified on a circle of radius $R$. We start with a discussion of the spectrum. 

\subsection{Spectrum}\label{spectrum}

As is well known, when $x$ is compactified on a circle, the corresponding left and right-moving momenta take the form  
\begin{equation}
	\label{compactk}
	p_L=\frac{n}{R}-\frac{wR}{\alpha'}~;\;\;\;\;\;p_R=\frac{n}{R}+\frac{wR}{\alpha'}~,
	\end{equation}
 where $(n,w)$ are the momentum and winding on the circle, respectively. The resulting spectrum is richer than in the non-compact case. Consider, for example,  the $(-1,-1)$ picture operators 
\begin{equation}
	\label{fullver}
	O_\Delta=\int d^2z e^{-\bar\varphi_+-\bar\varphi_-}e^{-iEt}e^{ip_Lx_L+ip_Rx_R}V_\Delta~,
	\end{equation}
where $V_\Delta$ is an arbitrary Virasoro primary of (left-moving worldsheet) dimension $\Delta$ in the Niemeier CFT $\M$. The mass-shell condition takes in this case the form  
\begin{equation}
	\label{compactmass}
	\frac{\alpha'}{4}E^2=\frac{\alpha'}{4}(p_R)^2=\frac{\alpha'}{4}(p_L)^2+\Delta-1 \ .
	\end{equation}
Plugging \eqref{compactk} into \eqref{compactmass}, gives the level matching condition 
\begin{equation}
	\label{levelm}
	\Delta-1=nw \ .
	\end{equation}
For $\Delta> 1$, \eqref{levelm} implies that $n$ and $w$ are both non-zero, and have the same sign. The energy $E$ \eqref{compactmass} takes the form 
\begin{equation}
	\label{massplus}
	E=\frac{n}{R}+\frac{wR}{\alpha'}~,\,\,\;\;\;{\rm for}\;\; n,w>0~,
	\end{equation}
and   
\begin{equation}
	\label{massminus}
	E=-\frac{n}{R}-\frac{wR}{\alpha'}~,\,\,\;\;\;{\rm for}\;\; n,w<0 \ .
	\end{equation}
One can think of these equations as follows. The winding term is a kind of zero point energy, the energy of a string of tension $T=1/2\pi\alpha'$ wrapped $|w|$ times around a circle of radius $R$, 
\begin{equation}
	\label{zeromass}
	E_0=\frac{|w|R}{\alpha'} \ . 
	\end{equation}
If we define the energy relative to $E_0$,  i.e. replace $E$ in \eqref{massplus}, \eqref{massminus} by $E_0+{\cal E}$, as is standard in the study of the dynamics on branes in string theory, we find 
\begin{equation}
	\label{eplus}
	{\cal E}=\frac{n}{R}~,\,\,\;\;\;{\rm for}\;\; w>0~,
	\end{equation}
and    
\begin{equation}
	\label{eminus}
	{\cal E}= -\frac{n}{R}~,\,\,\;\;\;{\rm for}\;\; w<0 \ .
	\end{equation}
Thus, for $w>0$ the spectrum only contains left-moving excitations on the wound string, while for $w<0$ it only contains right-moving ones. If one thinks of $w>0$ as describing wound fundamental strings, and $w<0$ as anti-strings, we find that the excitations on wound strings are left-moving, while those on anti-strings are right-moving.

Moreover, the energies \eqref{eplus}, \eqref{eminus} are related to the left-moving excitation level in the worldsheet Niemeier CFT $\M$. Indeed, according to \eqref{levelm}, we can write \eqref{eplus} as 
\begin{equation}
	\label{niem}
	R{\cal E}=n=\frac{\Delta-1}{w} \ .
	\end{equation}
For $w=1$, \eqref{niem} takes the form $R{\cal E}=\Delta-1$. This looks like the formula for the energy on the cylinder of a state with dimension $\Delta$ in a left-moving CFT with central charge $c=24$. It suggests that the spacetime theory describing $w=1$ states is closely related to the holomorphic Niemeier CFT that went into the construction of the $(0,2)$ heterotic string in question. We will return to this point later in the paper. Note also that for $\Delta=0$, i.e. $V_\Delta=1$, \eqref{niem} gives a negative energy, $R{\cal E}=-1$. Thus, superficially it looks like the energy is negative, however, for $R>l_s$ the total energy \eqref{massplus} is still positive, so one can use the above equations.

For $w>1$, equation \eqref{niem} has two notable properties. One is that it implies a constraint on $\Delta$, which is due to the fact that the momentum $n$ is integer. The second is that, as we will show in section \ref{symmprodn}, this formula is the same as that obtained in the $\mathbb{Z}_w$ twisted sector of the symmetric product of Niemeier CFT's. This hints that the spacetime theory for $w>0$ is closely related to such a symmetric product. Again, we will return to this point later. 

For $w=0$, \eqref{levelm} implies that only states with $\Delta=1$ exist, but they can have any momentum $n$. These states are described by the vertex operators $V_L^a$ \eqref{fullvL}, \eqref{fullv0}, and their right-moving counterparts $V_R^a$, with quantized spatial momentum $p=n/R$. As discussed in section \ref{noncomp}, the vertex operators \eqref{fullver} correspond in this case to pure gauge fields, see the discussion around \eqref{fullv0}, \eqref{aaa}. We will take the attitude explained there, that one should think of them not as states in the theory, but as symmetry generators that act on other states. We will study them from this point of view in the next subsection. 

The construction of the spectrum for $w\not=0$ described above is valid for operators $V_\Delta$ in the Niemeier CFT that are primary under worldsheet Virasoro. However, it can be generalized to all states in the Niemeier CFT. Consider, for example, a generalization of \eqref{fullver} where $V_\Delta$ is replaced by $\partial V_\Delta$, the derivative of a Virasoro primary operator in $\M$. This operator has worldsheet dimension $\Delta+1$, so the mass-shell condition for it is \eqref{compactmass}, with $\Delta\to\Delta+1$. Of course, we still have to satisfy level matching, \eqref{levelm}, which is not a problem for $|w|=1$, but places constraints on $\Delta$ for larger $|w|$. We will ignore these constraints in the discussion below. 

Even if the mass-shell condition is satisfied, the resulting operator is not BRST invariant. The reason is that the left-moving part of the operator, $\partial V_\Delta\exp\left(ip_Lx_L-iEt\right)$, is not primary under the worldsheet Virasoro $T(z)=T_x(z)+T_y(z)$, \eqref{leftxmu}, \eqref{leftyi}. Indeed, the operator $\partial V_\Delta$ satisfies the OPE
\begin{equation}
	\label{OPEpart}
	T_y(z)\partial V_\Delta(0)\sim {2\Delta V_\Delta(0)\over z^3}+{(\Delta+1)\partial V_\Delta(0)\over z^2}+{\partial^2 V_\Delta(0)\over z} \ .
	\end{equation}
The first term on the r.h.s. violates the BRST invariance of the operator constructed above. 

The solution of this problem is known. To cancel the triple pole term in \eqref{OPEpart}, we can consider the  operator
\begin{equation}
	\label{newver}
	e^{-\bar\varphi_+-\bar\varphi_-}\left(\partial V_\Delta-
 {2i\Delta\over wR}V_\Delta\partial x^-\right)e^{-iEt}e^{ip_Lx_L+ip_Rx_R}~,
	\end{equation}
 where $x^-$ is given in \eqref{xpm}.
One can check that this operator has the property that the triple pole in its OPE with the stress-tensor $T(z)$ cancels between the two terms in the brackets. Thus, integrating it over the worldsheet, as in \eqref{fullver}, gives a BRST invariant vertex operator. For $w=1$, this operator has energy larger than that of \eqref{fullver} by one unit, $R{\cal E}=n+1$.

Some comments about the preceding discussion are useful at this point:
\begin{itemize}
\item
The operator \eqref{newver} gives a BRST invariant observable corresponding to the operator $\partial V_\Delta$ in $\M$, for any Virasoro primary $V_\Delta$. One can generalize the discussion to {\it all} operators in $\M$. We will not discuss the general construction here. Thus, for $|w|=1$, the single-string spectrum looks like that of $\M$, while for $|w|>1$, it looks like that of the $Z_{|w|}$ twisted sector in the symmetric product of these CFT's (see section \ref{symmprodn}). 
\item In our construction above, we chose the improvement term (the second term in the brackets in \eqref{newver}) to only involve $\partial x^-$, and not $\partial x^+$, \eqref{xpm}. This is a gauge choice. We could modify \eqref{newver} by adding to it an arbitrary multiple of the operator 
\be e^{-\bar\varphi_+-\bar\varphi_-}\partial\left(V_\Delta e^{-iEt}e^{ip_Lx_L+ip_Rx_R}\right)~,\ee
which is BRST exact, and thus trivial. One advantage of the choice \eqref{newver} is that the coefficient of the improvement term only depends on the winding $w$, and not on the momentum $n$. In other words, it only depends on the winding sector, and not on the particular state within this sector. Any other choice would not have this property. This choice is also convenient for other reasons.
\item In the next subsection, we will describe a different way of constructing the full spectrum of the theory from the primaries \eqref{fullver}, by using a spectrum generating algebra related to an infinite-dimensional symmetry of the model.
\end{itemize}

\subsection{Symmetries I: Affine Lie algebra}\label{KM}

In this section, we will show that the $\N=2$ heterotic string has an infinite-dimensional symmetry algebra. In particular, if the Niemeier CFT $\M$ that enters the worldsheet construction has a current algebra $\widehat G$ \eqref{wsKM}, the spacetime theory contains the same affine Lie algebra. Recall that $\widehat G$ has rank twenty four, and is in general non-abelian (the Leech lattice is the only Niemeier lattice that gives an abelian affine Lie algebra $G=U(1)^{24}$). 

To construct this algebra, we go back to the analysis of sections \ref{noncomp}, \ref{spectrum}, and focus on the $w=0$ sector which, as we explained, is expected to contain symmetry generators. The level matching condition \eqref{levelm} implies that in that sector we must take $\Delta=1$, and there are two types of physical vertex operators, corresponding to left and right-movers in spacetime. We will first discuss the left-moving operators, and then comment on the right-moving ones. 

These operators are labeled by the integer momentum $n$. Generalizing \eqref{fullv0} to the compact case, it is natural to define   
\begin{equation}
	\label{defkan}
	K^a_n=\int d^2z J^a(z)\partial_{\bar z} e^{i{n\over R}x^-(z,\bar z)}~,\;\;\;\; n\in \mathbb{Z} \ .
	\end{equation}
Since the currents $J^a(z)$ are holomorphic on the worldsheet, the integrand of \eqref{defkan} can be written as a total derivative, at least away from other insertions in a worldsheet correlator. Thus, we can write 
\be K^a_n=\int d^2z \partial_{\bar z}\left(J^a e^{i{n\over R}x^-}\right)~,\ee
where the integral runs over the worldsheet, with small circles around the other insertions in the worldsheet path integral removed. Alternatively, we can write 
\begin{equation}
	\label{contour}
	K^a_n=\oint {dz\over 2\pi i} J^a(z) e^{i{n\over R}x^-(z,\bar z)}~,
	\end{equation}
where the contour integral runs counterclockwise around other insertions, and we slightly changed the overall normalization of the operator. We will discuss later some subtleties with the manipulations \eqref{defkan} -- \eqref{contour}, but for now we proceed. 

The commutator of two charges is given by
\begin{equation}
	\label{commK}
	[K^a_n, K^b_m]=\oint {d\xi\over 2\pi i}\oint_{\xi} {dz\over 2\pi i} J^a(z)e^{i{n\over R}x^-(z,\bar z)}J^b(\xi)e^{i{m\over R}x^-(\xi,\bar\xi)}\ .
	\end{equation}
The only contributions to the contour integral over $z$, going around $\xi$, are due to the poles in \eqref{wsKM}. The single-pole term gives $if^{ab}_{\,\,\,\,\,\,c} K^c_{n+m}$. The double pole gives 
\begin{equation}
	\label{central}
	\frac k2\delta^{ab}  {in\over R}\oint {d\xi\over 2\pi i}\partial x^- e^{i{n+m\over R} x^-}~.
	\end{equation}
For $n+m\not=0$, the operator \eqref{central} vanishes, since it is proportional to the integral of a total derivative of a ``good'' operator in the theory, $e^{i{n+m\over R}x^-}$. However, for $n+m=0$, the situation is different. In that case, the operator \eqref{central} is a derivative of the operator $x^-$, which is {\it not} a good operator in the theory, and thus, it does not need to decouple. 

Combining both contributions to the commutator \eqref{commK}, we find 
\begin{equation}
	\label{commutrel}
	[K^a_n, K^b_m]=if^{ab}_{\,\,\,\,\,\,c}K^c_{n+m}+
 \frac k2 n\delta^{ab}\delta_{n+m,0}P^-_L~,
	\end{equation}
where 
\begin{equation}
	\label{plm}
	P_L^-=\frac1R\oint\frac{dz}{2\pi i} i\partial x^- \ .
	\end{equation}
We see that the operators $K^a_n$ form an affine Lie algebra $\widehat G$ of level $k P_L^-$. Interestingly, the central extension is an operator, rather than a scalar, but it commutes with all the $K^a_n$. Thus, it is central, as expected. We will provide a better understanding of the operator \eqref{plm}, and its role in \eqref{commutrel},  later. Note also that the construction of an affine Lie algebra \eqref{contour} -- \eqref{plm} is directly related to the current algebra discussed in section \ref{noncomp}, around equation \eqref{curalg} -- \eqref{curralgebra}. 

The above discussion can be repeated for the right-moving symmetry generators, defined as in \eqref{defkan}, \eqref{contour}. They are given by 
\begin{equation}
	\label{contourplus}
	\bar K^a_m=\oint {dz\over 2\pi i} J^a(z)e^{i{m\over R}x^+(z,\bar z)} \ .
	\end{equation}
A similar calculation to \eqref{commK} -- \eqref{plm} shows that $\bar K^a_m$ satisfy a right-moving $\widehat G$ affine Lie algebra with central extension $kP_L^+$, where $P_L^+$ is given by \eqref{plm} with $-\to +$. 

We note in passing that the charges $K^a_0$, \eqref{contour}, and $\bar K^a_0$, \eqref{contourplus}, are the same. In other words, this model has a single global symmetry, $G$, that acts both on the left-movers and on the right-movers. This should be contrasted with CFT's with left and right Kac-Moody symmetries, $\widehat G_L$, $\widehat G_R$, which have 
separate global symmetries $G_L$ and $G_R$. This suggests that the spacetime theory is not conformal. We already saw a hint of that in section \ref{noncomp}, where we wrote the spacetime action \eqref{NGact}, which of course is not conformal. We will return to this issue below. 

We now reach an important subtlety in the above discussion. Suppose we want to generalize the calculation \eqref{commK} -- \eqref{plm}, to compute the commutator of left and right-moving symmetry generators, $[K^a_n, \bar K^b_m]$. The analog of \eqref{commK} contains in this case the OPE $e^{i{n\over R}x^-(z,\bar z)}e^{i{m\over R}x^+(\xi,\bar \xi)}$, which goes like $|z-\xi|^\beta$, with $\beta$ proportional to $nm\alpha'/R^2$. In evaluating the above commutator we are led to consider integrals such as 
\be
\label{badact}
\oint_{\xi} {dz\over z-\xi}|z-\xi|^\beta \ . \ee
This presents two problems. One is that the integral \eqref{badact} is, in general, divergent. The second is that since it is an integral of a non-holomorphic function, the result depends on the shape of the integration contour. 

We learn that correlators with insertions of {\it both} \eqref{contour} {\it and} \eqref{contourplus} are more subtle, for general $n,m$. At first sight, it seems that this means that these symmetry generators  should be discarded (except for the zero modes $K^a_0=\bar K^a_0$). To see that the situation is actually better, it is useful to consider the action of the charges \eqref{contour}, \eqref{contourplus} on states with $w\not=0$. 

Since sectors with $w>0$ only contain left-moving excitations, while those with $w<0$ contain only right-moving ones (see the discussion around equations \eqref{eplus}, \eqref{eminus}), it is natural to expect that the left-moving symmetry generators $K^a_n$ act well on the sectors with $w>0$, and the $\bar K^a_m$ act on the sectors with $w<0$. We next show that this is indeed the case. 

Suppose the physical vertex operator \eqref{fullver} has the property that the worldsheet operator $V_\Delta$ that enters its construction is not only a primary of Virasoro, but is also a primary of the $\widehat G$ affine Lie algebra \eqref{wsKM}. Thus, it satisfies the worldsheet OPE
\begin{equation}
	\label{repR}
	J^a(z)V_\Delta(\xi) \sim \frac {T^a V_\Delta(\xi)}{z-\xi}~,
	\end{equation}
where $T^a$ are matrices representing the Lie algebra $G$ in a representation $R$. In other words, the numerator on the r.h.s. of \eqref{repR} looks like $(T^a)^i_jV_\Delta^j$, with $i,j=1,\cdots, {\rm dim}\; R$ (summed over~$j$), and  $V_\Delta$ on the l.h.s. needs to be replaced by $V^i_\Delta$.

As explained in subsection \ref{spectrum}, the operators \eqref{fullver} describe left or right-moving excitations, depending on the sign of the winding $w$. For $w>0$, they are left-moving, and according to the discussion above, we expect the affine Lie algebra \eqref{contour} to act on them. 

To see that action, we compute the commutator
\begin{equation}
	\label{commVR}
	[K^a_l, O^i_\Delta]=\oint \frac{dz}{2\pi i} J^a(z) e^{i\frac lRx^-(z,\bar z)}e^{-\bar\varphi_+-\bar\varphi_-}e^{-iEt}e^{ip_Lx_L+ip_Rx_R}V^i_\Delta(\xi)~,
	\end{equation}	
where $p_L$, $p_R$ and $E$ are given by \eqref{compactk}, \eqref{massplus}, and we omitted the $\xi$ integral on the r.h.s.

Consider first the case $l=0$. In this case, the only singularity as $z\to\xi$ is the one in \eqref{repR}, and one gets 
\be 
[K^a_0, O^i_\Delta]=(T^a)^i_jO^j_\Delta \ . \ee
This is the statement that the vertex operator  \eqref{fullver} transforms in the representation~$R$ of~$G$. 

For general $l$ in \eqref{commVR}, a short calculation gives an additional factor $(z-\xi)^{lw}$.\footnote{Note that there is no dependence on $\bar z-\bar\xi$, despite the fact that the operators we are contracting, $x^\pm$, \eqref{xpm}, are not holomorphic.} The additional power of $z-\xi$ has the following effect on the commutator \eqref{commVR}. For $l>0$, it gives 
\begin{equation}
	\label{lpositive}
	[K^a_l, O^i_\Delta]=0\;\;\;{\rm for}\;\;\; l>0 \ .
	\end{equation}	
This is natural, since $K^a_l$ are in this case annihilation operators, and the state corresponding to the vertex operator $O^i_\Delta$ is primary. For $l<0$, the situation is more interesting. The leading singularity in the OPE \eqref{commVR} is $1/(z-\xi)^{1+|l|w}$, and in order to compute the contour integral we need to expand the integrand to $|l|w$'th order and pick out the single pole term. 

The resulting operator \eqref{commVR} has the following properties:
\begin{itemize}
\item It is BRST invariant, since it is obtained by commuting two BRST invariant operators, $K^a_l$ and $O^i_\Delta$. 
\item If the original operator $O^i_\Delta$ has momentum and winding $(n,w)$, the operator \eqref{commVR} has $(n+|l|,w)$. In other words, it is an operator that lives in the same winding sector, and has energy larger by $|l|$ units from the original one. This is consistent with our expectation that $K^a_l$ with negative $l$ are raising operators in $\widehat G$ representations, that increase the energy by $|l|$.
\item Since the operator \eqref{commVR} is obtained by expanding the r.h.s. to order $|l|w$, one can think of it as having $\Delta$ larger by that amount from the original operator \eqref{fullver}. This is compatible with the level matching condition \eqref{levelm}. If the original operator \eqref{fullver} satisfies level matching, so does the new one, since it is obtained by taking $n\to n+|l|$, $\Delta\to \Delta+|l|w$. 
\end{itemize}

One can summarize the above discussion by saying that the operators $K^a_l$ with negative~$l$ form a spectrum generating algebra for $\widehat G$. Starting with primaries of $\widehat G$, \eqref{fullver}, \eqref{repR}, we can act with the raising operators, $K^a_l$ with $l<0$, and construct all the descendants, in a sector of given $w$. Of course, if one acts with $K^a_l$ with $l>0$ on the resulting $\widehat G$ descendants, one does not get zero, as in \eqref{lpositive}, but rather another state in the $\widehat G$ representation. This is guaranteed by the algebra \eqref{commutrel}. 

It is also instructive to calculate the central extension of the spacetime $\widehat G$ algebra \eqref{commutrel}. For this we need to calculate the commutator of $P_L^-$, \eqref{plm}, with $O_\Delta$ \eqref{fullver}. A short calculation gives 
\begin{equation}
	\label{plw}
	[P_L^-,O_\Delta]=w O_\Delta \ .
	\end{equation}
Thus, we conclude that in a sector with given $w$, the central extension in \eqref{commutrel} is a constant, and the level of the affine Lie algebra $\widehat G$ is $kw$, $w$ times the worldsheet level, which in our case is equal to one. The operators $K^a_l$ constructed above do not change $w$, which is consistent with the fact that the central extension is fixed in a given representation of $\widehat G$.

Above, we discussed the physical states in sectors with $w>0$, and studied the action of the left-moving Kac-Moody currents $K^a_n$ \eqref{contour} on them. One can repeat the discussion for sectors with $w<0$, which are acted on by the right-moving Kac-Moody currents $\bar K^a_n$, \eqref{contourplus}. The details are very similar, and we will not repeat them. 

Like in our discussion of the commutator of $K^a_n$ and $\bar K^b_m$ around equation \eqref{badact}, the action of $K^a_n$ on the right-moving vertex operators, the analog of \eqref{commVR} with $w<0$, runs into the same problems as there. Similarly, the right-moving modes $\bar K^a_n$ do not act well on the left-moving operators, \eqref{fullver} with $w>0$. 

One can think of the situation in the following way. In addition to the fact that $K^a_n$~$(\bar K^a_n)$ give rise to a spectrum generating algebra for $w>0$ $(w<0)$, they can be used inside a subset of correlation functions. In particular, the operators $K^a_n$ can be inserted into correlation functions of the form 
\be
\label{corrfn}
\langle O_{\Delta_1}\cdots O_{\Delta_m}\rangle~,\ee
where all the $O$'s are either the operators \eqref{fullver} with $w_j>0$ (for the in-states), or their c.c.'s (for the out-states). Similarly, the operators $\bar K^a_n$ can be inserted into correlation functions of the form \eqref{corrfn}, where all the $O$'s have $w_j<0$. Correlation functions that contain both operators with $w>0$ and $w<0$ do not transform well under either Kac-Moody algebra. 

As we will discuss in section \ref{spdyn}, the above properties of the left and right-moving affine Lie algebras can be understood from the spacetime point of view as due to a coupling of left and right-movers via $g_s$ effects. For now, we note that this issue (the breaking of the Kac-Moody algebras in general string amplitudes) afflicts $n$-point functions with $n>2$, so it is in a sense a $g_s$ correction to the free string theory, which is the focus of this section.

\subsection{Symmetries II: Virasoro}
\label{Vir}

Given the construction of the spacetime affine Lie algebra in subsection \ref{KM}, it is natural to ask whether the $(0,2)$ heterotic string also has a Virasoro algebra. This is the problem we will address in this subsection. Before turning to that discussion, we note that for any affine Lie algebra, one can construct a corresponding Sugawara stress-tensor. In particular, one can do this for the affine Lie algebra $\widehat G$ \eqref{contour}. As we will explain later, this is {\it not} what we are looking for.  

The construction of the $K^a_n$ suggests that the Virasoro generators $L_n$  should be in the $w=0$ sector as well, and should go like $e^{i{n\over R} x^-}$. Two natural candidates are
\begin{equation}
	\label{candln}
	\oint \frac{dz}{2\pi i} \partial x^\pm e^{i\frac nR x^-} \ .
	\end{equation}
The operator which contains $\partial x^-$ was discussed before, around equation \eqref{central}. We saw that for $n\not=0$ it is trivial, but it may play a role in the construction of $L_0$. We will see later that it indeed does. 

For $n\not=0$, we are left with the operator that contains $\partial x^+$ in \eqref{candln}. The problem with this operator is that it is not BRST invariant. This is due to the fact that the operator $\partial x^+ e^{i\frac nR x^-}$ is not primary under the worldsheet Virasoro \eqref{leftxmu}. Indeed, the OPE $T_x(z)\partial x^+ e^{i\frac nR x^-}(0)$  contains a cubic pole, whose coefficient is proportional to   $n e^{i\frac nR x^-}(0)$ (see appendix \ref{apenvir}).

In order to make a physical operator out of \eqref{candln}, we need to add to it another operator that cancels this cubic pole. Interestingly, the authors of~\cite{Callebaut:2019omt} (see also~\cite{Baba:2009ns,Biswas:2024unn}) encountered a similar problem in a different context, and proposed a solution for it. Phrasing it in our language, and restoring the numerical coefficients, their proposal is to consider the operators  
\begin{equation}
	\label{Ln}
	L_n=\oint\frac{dz}{2\pi i}\left({iR\over\alpha'}\partial x^++\frac n2\partial\ln(\partial x^-)\right)e^{i\frac nR x^-(z,\bar z)} \ .
	\end{equation}
The second term in \eqref{Ln} is non-local, and at first sight appears to be problematic. We will return to this issue below, but for now we assume that it is well defined and proceed. 

The first question is whether \eqref{Ln} is BRST invariant. It is not hard to see that the cubic pole in the OPE of the worldsheet stress-tensor $T(z)$ with the integrand of \eqref{Ln} cancels between the two terms in the brackets (see appendix \ref{apenvir}). Thus, $L_n$ is indeed physical. The second question is why we did not see the vertex operator \eqref{Ln} in our discussion of the spectrum in the $w=0$ spectrum in subsection \ref{spectrum}. The answer is that this vertex operator is non-local on the worldsheet, and is therefore not captured by the standard analysis. We believe it should still be included, since the $w=0$ sector describes symmetry generators rather than states, and we see no obstruction to including \eqref{Ln} in the list of such generators.

The next question concerns the commutator $[L_n, L_m]$. A calculation, described in appendix \ref{apenvir}, gives 
\begin{equation}
	\label{viralgggg}
	[L_n,L_m]=(n-m)L_{n+m}+{C\over 12}n^3\delta_{n+m,0}~,
	\end{equation}
where $C$ is given by 
\begin{equation}
	\label{formC}
	C=24P^-_L~,
	\end{equation}
with $P^-_L$ \eqref{plm}.

The algebra \eqref{viralgggg} is not quite the Virasoro algebra, since the central term goes like $n^3$, and not like $n^3-n$. However, this is easy to fix. If we set $n+m=0$ in \eqref{viralgggg}, the r.h.s. looks like 
\begin{equation}
	\label{lzero}
	2nL_0+{C\over 12}n^3 \ .
	\end{equation}
If we shift $L_0$ by $C/24$, i.e. write 
\begin{equation}
	\label{formL0}
	L_0=\oint\frac{dz}{2\pi i}\left({iR\over\alpha'}\partial x^++{i\over R}\partial x^-\right)~,
	\end{equation}
the algebra \eqref{viralgggg} becomes the Virasoro algebra,
\begin{equation}
	\label{viralg}
	[L_n,L_m]=(n-m)L_{n+m}+{C\over 12}(n^3-n)\delta_{n+m,0} \ .
	\end{equation}
This shift has a natural interpretation. It means that $L_0$ in \eqref{Ln} is actually ($R$ times) the energy on the cylinder, and the shift is the usual one that relates the two. As we saw before, the central charge $C$, \eqref{formC}, is given in the sector with winding $w$ by $C=24w$, see \eqref{plw}. We will provide an interpretation of this result later in the paper. 

Now that we constructed the Kac-Moody and Virasoro operators \eqref{contour}, \eqref{Ln}, \eqref{formL0}, we can calculate the commutator of the two. We find
\be [L_n,K^a_m]=-mK^a_{n+m}~,\ee
which is the standard commutation relation between Virasoro and Kac-Moody generators. 

Finally, we can generalize the discussion of the action of Kac-Moody generators on states in sectors with $w>0$ (the discussion around equation \eqref{commVR}) to the Virasoro case. For $L_0$, \eqref{formL0}, we find  
\begin{equation}
	\label{commL0}
	[L_0, O_\Delta]=(n+w)O_\Delta~,
	\end{equation}
where $O_\Delta$ is given by \eqref{fullver}. This equation has a natural interpretation: the $w$ term is the value of $C/24$ in the sector with winding $w$. Therefore, the energy on the cylinder of the state corresponding to $O_\Delta$ is 
\begin{equation}
	\label{energyw}
	RE_\Delta=L_0-C/24=n={\Delta-1\over w}~,
	\end{equation}
where the last equality follows from \eqref{levelm}. We will interpret \eqref{energyw} in terms of symmetric product CFT later in the paper. Note that \eqref{energyw} agrees with the energy above the vacuum \eqref{eplus}, that we computed above.

We next turn to the calculation of $[L_l,O_\Delta]$ for $l\not=0$. In this case, we have to face the fact that the operator \eqref{Ln} is non-local; in particular, writing 
\be 
\label{lndef}
\partial\ln(\partial x^-)={\partial^2 x^-\over \partial x^-} \ , \ee
we have to understand how to deal with the inverse of $\partial x^-$ that appears in this expression. 

First, we note that the operators \eqref{fullver} depend on $x_L^+$, the conjugate variable to $\partial x^-$, only through the factor 
\be  e^{ip_L^-x_L^+}=e^{-{iwR\over\alpha'} x_L^+} \ . \ee
Thus, to evaluate the above commutator, we need the OPE of an arbitrary function of $\partial x^-$,  ${\cal F}(\partial x^-(z))$, with this factor. The authors of~\cite{Callebaut:2019omt} show how to perform this calculation. In our notation, their result is 
\begin{equation}
	\label{fxminus}
{\cal F}(\partial x^-(z))e^{-{iwR\over\alpha'} x_L^+(0)}\sim
\left[{\cal F}\left(-{iwR\over z}+\partial x^-(z)\right)-{\cal F}(\partial x^-(z))\right] e^{-{iwR\over\alpha'} x_L^+(0)}~.
	\end{equation}
Using this result for the function $\cal F$ that appears in \eqref{Ln}, 
\begin{equation}
	\label{formF}
{\cal F}(\partial x^-(z))=\partial\ln\partial x^-~,
	\end{equation}
we find the OPE 
\begin{equation}
	\label{OPEF}
{\cal F}(\partial x^-(z))e^{-{iwR\over\alpha'} x_L^+(0)}\sim-{1\over z}e^{-{iwR\over\alpha'} x_L^+(0)} \ .
	\end{equation}
We can use the OPE \eqref{OPEF} to evaluate the commutator $[L_l,O_\Delta]$. 

As in the Kac-Moody calculation of subsection \ref{KM}, there is a difference between positive and negative $l$. For positive $l$, we again get zero, since the contour integral \eqref{Ln} going around the location of \eqref{fullver} behaves as 
\be  \oint_0 dz z^{lw-1}(\cdots)~,\ee 
which vanishes for positive $l,w$. For $l<0$, we again have an order $|l|w+1$ pole, and have to expand the integrand to order $|l|w$ in $z$ to find the single-pole contribution. The interpretation is the same as in the Kac-Moody analysis. The Virasoro generators $L_l$ with $l>0$ are annihilation operators which, when acting on the Virasoro primaries described by \eqref{fullver}, give zero. The Virasoro generators with negative $l$ are spectrum generating operators, which map primaries of the Virasoro algebra \eqref{viralg} to descendants. 

So far we discussed the left-moving Virasoro generators \eqref{Ln}, \eqref{formL0}, and their action on the left-moving vertex operators with $w>0$. There is of course also a right-moving Virasoro algebra with generators $\bar L_n$, which is obtained from \eqref{Ln}, \eqref{formL0} by exchanging $x^+$ with $x^-$ in these equations. The generators of this algebra act naturally on states with $w<0$. As in the Kac-Moody case, the left-moving Virasoro generators don't act well on right-moving excitations (including on right-moving Virasoro generators), and vice-versa.

\subsection{Symmetries III: Higher spin symmetries}
\label{HS}

In the previous subsections, we constructed the generators of spin one and two symmetries that act naturally on chiral excitations of the $(0,2)$ string. Of course, all operators in the Niemeier CFT $\M$ can be viewed as symmetry generators, since they are holomorphic, and form the chiral algebra of the model. Thus, we should be able to generalize the discussion above to all operators in $\M$.

A large class of such operators consists of the Virasoro primaries $V_\Delta$ defined in \eqref{fullver}. Following the discussion of this section, we expect to be able to associate with each $V_\Delta$ an infinite set of modes, $V_n$, that satisfy an infinite-dimensional algebra, and act on the chiral excitations in sectors with $w>0$.

Following the discussion of the Virasoro generators above, it is natural to write 
\begin{equation}
	\label{higherS}
	V_n=\oint \frac{dz}{2\pi i} 
 {V_\Delta\over (\partial x^-)^{\Delta-1}}
 e^{i\frac nR x^-}\ .
	\end{equation}
For $\Delta=0$, i.e. $V_\Delta=1$, \eqref{higherS} vanishes for $n\not=0$, as explained around eq. \eqref{central}, while for $n=0$ it reduces to \eqref{plm}. For $\Delta=1$, i.e. $V_\Delta=J^a$ (which are defined in section \ref{noncomp}, around eq. \eqref{wsKM}), it gives \eqref{contour}. For $\Delta>1$, one has to make sense of the operator $(\partial x^-)^l$ with negative $l$. We can define it by continuation from non-negative $l$. In particular, since $(\partial x^-)^le^{i\frac nR x^-}$ is a Virasoro primary of dimension $l$ for any $l\in\mathbb{Z}_+$, it formally has this property for $l<0$ as well. Therefore, \eqref{higherS} is BRST invariant. 

We note in passing that one can think of \eqref{higherS} in the following way. As mentioned above (in footnote 4), in correlation functions of operators of the form \eqref{fullver} with $w>0$, the worldsheet field $x^-(z,\bar z)$ is holomorphic, $\bar\partial x^-=0$. Therefore, one can trade the contour integral over $z$ in \eqref{higherS} for one over $x^-(z)$. Plugging the transformation property of $V_\Delta$ under this transformation, 
\be
\label{vxz}
V_\Delta(z) (dz)^\Delta=V_\Delta(x^-) (dx^-)^\Delta~,
\ee
into \eqref{higherS}, we find that it can be written as 
\begin{equation}
	\label{higherSS}
	V_n=\oint \frac{dx^-}{2\pi i} 
 V_\Delta(x^-)
 e^{i\frac nR x^-}\ .
	\end{equation}
The variable $x^-$ lives on a cylinder, and the integral in \eqref{higherSS} is around the spatial circle. We see that $V_n$ can be thought of as the mode with momentum $n/R$ of the field $V_\Delta$ on the spacetime cylinder.

To use \eqref{higherS} in calculations, we can, for example, compute the commutation relations of $V_n$ with $K^a_m$. If $V_\Delta$ is a primary of the worldsheet Kac-Moody algebra $\widehat G$ in the representation $R$, \eqref{repR}, one finds
\be  [K^a_n, V^i_m]=T^a_{ij}V^j_{n+m} \ . \ee
One can also compute the commutation relation between different $V_n$, and the action of $V_n$ on the states in sectors with $w>0$, \eqref{fullver}. This was done in the follow-up paper \cite{Giveon:2024vlh}. 

Finally, we note that the construction of higher spin fields \eqref{higherS}, which was done for Virasoro primaries, can be generalized to descendants, but the structure of the resulting operators is more complicated. We already saw an example of this phenomenon in the construction of the stress tensor in the previous subsection. The analog of \eqref{higherS} for that case would naturally involve the replacement of $V_\Delta$ by $T_y(z)$, the stress tensor of the worldsheet Niemeier CFT $\M$, \eqref{leftyi}. However, since $T_y$ is not primary, we need to correct this structure. This raises two questions: 
\begin{itemize}
\item How to correct \eqref{higherS} for this case ($\Delta=2$, $V_\Delta=T_y$) to make it BRST invariant.
\item We constructed the spacetime stress-tensor in the previous subsection, see \eqref{Ln}, \eqref{formL0}. The construction of this subsection should give an equivalent set of Virasoro generators $L_n$. Why is this the case? 
\end{itemize}
To address the first question, we follow the procedure outlined around \eqref{higherS} -- \eqref{higherSS}. In the spacetime theory, we can define the operators 
\begin{equation}
	\label{higherSSS}
	L_n=-iR\oint \frac{dx^-}{2\pi i} 
 T_y(x^-)
 e^{i\frac nR x^-}\ ,
	\end{equation}
which are modes of the spacetime stress tensor in the left-moving spacetime Niemeier CFT. As in \eqref{vxz}, we want to pull back \eqref{higherSSS} to the worldsheet, using the conformal map $x^-=x^-(z)$. For the stress tensor, we need to take into account the Schwarzian derivative term,
\be 
\label{schwarz}
(\partial x^-)^2T_y(x^-)=T_y(z)-2\{x^-,z\}~,\;\;\;
\{x^-,z\}={\partial^3 x^-\over \partial x^-}-{3\over2}{(\partial^2 x^-)^2\over (\partial x^-)^2}~.\ee 
Plugging \eqref{schwarz} into \eqref{higherSSS}, we find 
\begin{equation}
\label{newLn}
L_n=-iR\oint \frac{dz}{2\pi i} 
\left[ {T_y(z)\over \partial x^-}
-2{\partial^3 x^-\over(\partial x^-)^2}+3{(\partial^2 x^-)^2\over(\partial x^-)^3}
 \right]e^{i\frac nR x^-}\ .
	\end{equation}
One can check that the operator \eqref{newLn} is a worldsheet Virasoro primary, and thus is physical. Moreover, it satisfies\footnote{See appendix A of \cite{Eberhardt:2019qcl} for a similar calculation in a different context.} the Virasoro algebra \eqref{viralgggg}. Like in the discussion around eq.~\eqref{lzero}, we need to shift $L_0$ by $C/24$ to get the algebra \eqref{viralg}. This shift has the same origin as there -- it is due to the fact that \eqref{higherSSS} with $n=0$ is the energy on the cylinder.

The stress tensor \eqref{newLn} looks superficially different from the one we constructed in the previous subsection, \eqref{Ln}, \eqref{formL0}. In appendix \ref{apenvirrr} we show that, as expected, they are the same in the BRST cohomology. One can think of  \eqref{newLn} as the ``transverse'' form, built from $T_y$, and of \eqref{Ln} as the ``longitudinal'' form, built from $\partial x^+$. They are related by the Virasoro constraint $T_x+T_y\simeq0$, in the spirit of DDF versus light-cone operators. 

The construction of equations \eqref{higherSSS} -- \eqref{newLn} can be generalized to an arbitrary descendant operator $V_\Delta$ in the Niemeier CFT. To do this, one starts with the Fourier transform \eqref{higherSS}, and uses the analog of \eqref{schwarz} for $V_\Delta$ (or, equivalently, the OPE of $V_\Delta$ with the stress tensor), to arrive at the analog of \eqref{newLn}. We will not present the details of this construction here.

\section{Thermal partition sum} \label{torus}

In this section, we compute the (spacetime) thermal partition sum for the $(0,2)$ string, as a check on the calculation of the spectrum in section \ref{comp}. This partition sum is defined as 
\be Z(\beta,\mu, b) = \mbox{Tr}\, e^{-\beta \left(H  - R b w  / \alpha'+ i \mu n / R\right)} \ , \label{zthermal} \ee
where the trace runs over all single-string states, and $\beta$, $b$, and $\mu$ are chemical potentials that couple to the energy, winding and momentum, respectively. 

From the worldsheet point of view, to compute \eqref{zthermal} we need to evaluate the torus partition sum in the $(0,2)$ string theory \cite{Polchinski:1985zf}, with the target space being a Euclidean torus of size $2\pi R$ and modulus 
\be \zeta ={(\mu + i) \beta \over 2\pi R}~. \label{zetaa} \ee
Parametrizing the target-space torus by the complex coordinate $x$, with metric $ds^2=dxd\bar x$, this corresponds to the identifications
\be x \sim x + 2 \pi R \sim x+ 2 \pi R \zeta~. \ee
The chemical potential for winding, $b$ in \eqref{zthermal}, corresponds to turning on an imaginary $B$-field $ib$ on the torus (so that the $B$-field is real in the Lorentzian continuation). It is normalized such that for $b=1$ the winding term in \eqref{zthermal} subtracts from $H$ the energy due to winding (\eqref{zeromass} for $w>0$).  

The torus partition sum for the $(0,2)$ string on a target-space torus is given by (see e.g. appendix A of \cite{Kutasov:1996vh})
\be Z = \int_{\cal F} {d\tau d \bar \tau \over \tau_2^2}  Z_0(\tau, \bar \tau) Z_{\M}(\tau)~.\label{partsum} \ee
Here, $\tau=\tau_1+i\tau_2$ is the modulus of the worldsheet torus,  ${\cal F}$ is the fundamental domain over which it is integrated, $Z_0(\tau,\bar\tau)$ is the contribution to the partition sum of the spacetime fields $x^\mu$ and the ghosts, and $Z_{\M}(\tau)$ is the partition sum of the Niemeier CFT, $\M$, that enters the worldsheet construction (see section \ref{noncomp}). The contributions of the non-zero modes of $x^\mu$ and the ghosts to $Z_0$ cancel, and one is left with the contribution of the zero modes,
\be  Z_0(\tau, \bar \tau)= \tau_2 \sum_{{m_1, m_2, \atop n_1, n_2}}\exp \left(- 2 \pi \tau_2 {\bf H} + 2 \pi i \tau_1 {\bf P}\right)~, \label{zeromode}\ee
with (see (2.4.11) of \cite{Giveon:1994fu} and (3.55) of \cite{Chakraborty:2024mls})
\be\label{HHH}
{\bf H} = {1\over 2}\left[n_i (g^{-1})^{ij} n_j + m^i(g - b g^{-1} b)_{ij} m^j + 2 m^i b_{ik} (g^{-1})^{kj} n_j\right]~, \;\;\; {\bf P} = m^in_i~,\;\;\; i,j=1,2\ , \ee
and
\be\label{ggg} g_{ij} = {R^2 \over \alpha'} \left(\begin{array}{cc}1 &{\zeta_1} \\ {\zeta_1}& {|\zeta|^2} \end{array}\right)~, \qquad b_{ij}=ib{R^2\over\alpha'} \left(\begin{array}{cc}0 & -\zeta_2 \\ \zeta_2 & 0 \end{array}\right)~, 
\ee
where $n_i$ is the momentum quantum number along the $i$-th coordinate, and $m^i$ is the winding quantum number.

Plugging \eqref{zeromode} into \eqref{partsum}, we find
\be Z = \sum_{{m_1, m_2, \atop n_1, n_2}}\int_{{\cal F}} {d \tau d \bar \tau \over \tau_2}e^{ - 2 \pi \tau_2 {\bf H}  + 2 \pi i \tau_1 {\bf P} } Z_{\cal M}(\tau)~. \label{Zheta}\ee
This partition function (\ref{Zheta}) can be rewritten in a form that makes the spectrum more manifest, by using the unfolding trick described in \cite{Polchinski:1985zf}. 
This rewriting involves Poisson resumming $n_2$ as a sum of $w_2$. A useful way of thinking about this resummation is as a  Fourier transform,
\be \sum_{n_2} f(n_2) = \int dn_2\,  f(n_2)  \sum_{w_2}\delta(n_2 -w_2) =    \int dn_2 \, f(n_2) 
 \sum_{w_2} e^{2 \pi i n_2 w_2} = \sum_{w_2} \tilde f(w_2) \ . \label{poisson} \ee
Plugging \eqref{poisson} into \eqref{Zheta}, we have
\be \label{zzzz}
Z = \sum_{{m_1, m_2, \atop n_1, w_2}} \int_{{\cal F}} {d \tau d \bar \tau \over \tau_2}  \int d n_2 \, e^{- 2 \pi \tau_2 {\bf H}  + 2 \pi i \tau_1 {\bf P} + 2 \pi i n_2 w_2} Z_{\cal M}(\tau)~. \ee
After summing over $m_1$, $n_1$, and integrating over $n_2$, the integrand of the $d \tau d \bar \tau/\tau_2^2$ integral in \eqref{zzzz} is invariant under the $SL(2,\mathbb{Z})$ which acts on the worldsheet modulus  $\tau$, provided that $(w_2,m_2)$ transforms as a doublet.  Thus, we can trade the sum over $w_2, m_2\in \mathbb{Z}$ and a $\tau$ integral over the fundamental domain ${\cal F}$, \eqref{zzzz}, for a sum over $w_2=1,2,\dots$, with $m_2=0$, and a $\tau$ integral over the strip ${\cal S}$ corresponding to  $|\tau_1|\le {1\over 2}$, $\-\infty<\tau_2<\infty$,
\be \label{zzz}
Z = \left. \sum^\infty_{w_2=1} \sum_{m_1,  n_1 }  \int_{{\cal S}} {d \tau d \bar \tau \over \tau_2}   \int dn_2\,     e^{- 2 \pi \tau_2 {\bf H}  + 2 \pi i \tau_1 {\bf P} + 2 \pi i n_2 w_2 }\right|_{m_2=0} Z_{{\cal M}}(\tau)~. \ee
After performing the Gaussian integral over $n_2$, the sum over $w_2$ can be interpreted as a sum over the number of strings in a multi-string gas \cite{Polchinski:1985zf}. Writing \eqref{zzz} as
\be  Z = \sum_{w_2=1}^\infty Z_{w_2}~,  \ee
the single-string contribution is
\be \label{singles}
Z_{w_2=1}  = \sum_{n_1, m_1} \int_{{\cal S}} {d \tau d \bar \tau \over \tau_2^{3/2}}e^{-2 \pi \tau_2\left( {\alpha'  \over  2R^2} n_1^2 +  {R^2 \over 2 \alpha'} m_1^2 \right)  
 + 2 \pi i \tau_1 m_1 n_1  -  2 \pi i \zeta_1 n_1  - {2 \pi R^2 \zeta_2 b m_1 \over \alpha'} -{\pi R^2 \zeta_2^2 \over  \alpha' \tau_2} } Z_{\cal M}(\tau)~.
\ee
To read off the spectrum from the partition function  \eqref{Zheta}, \eqref{singles}, we write
\be Z_{\M}(\tau) ={\mbox Tr}\, q^{L_0-c/24}= \sum_{\Delta} D(\Delta) e^{-2 \pi \tau_2 (\Delta - 1) + 2 \pi i \tau_1 (\Delta - 1)}~, \label{ZNiemeier}\ee
where in the second equality we used $q=e^{2\pi i\tau}=e^{2\pi i\tau_1-2\pi\tau_2}$, and $c=24$. Here, $\Delta\in \mathbb{Z}_+$ are the left-moving dimensions of the operators in $\M$, and $D(\Delta)$ are their degeneracies.

Plugging \eqref{ZNiemeier} into \eqref{singles}, the integral over $\tau_1$ and $\tau_2$ can be performed in closed form. A useful identity for the $\tau_2$ integral is
\be  \int_0^\infty {dt \over t^{3/2}} \,  e^{-a/t -ct} = \sqrt{\pi \over a} e^{-2 \sqrt{ac}}  \ . \ee
The result is 
\be Z_{w_2=1} =\sum_{n_1, m_1} \sum_\Delta  D(\Delta) e^{-2 \pi \zeta_2 R ( \sqrt{({ n_1^2 / R^2}) + (R^2 / \alpha'^2) m_1^2 + {(2 / \alpha')} (\Delta -1)}+ R b m_1/\alpha') - 2 \pi i \zeta_1 n_1} \delta_{ m_1 n_1 + \Delta ,1}~.\label{z11a}\ee
The Kronecker delta constraint in \eqref{z11a} arises as a result of integrating over $\tau_1\in [-{1\over2},{1\over2}]$.

Comparing \eqref{z11a} to \eqref{zthermal}, \eqref{zetaa}, we see that the integers $(n_1, m_1)$ correspond to momentum and winding on $\mathbb{S}^1$, respectively, and the spectrum we read off \eqref{z11a} is the same as that obtained in section \ref{comp}. Indeed, setting $m_1 = - w < 0$, and $n_1= n$, we obtain
\be Z_{w_2=1}^{(w)} =\sum_{n \ge 0}  D(1+nw)   e^{-2 \pi \zeta_2 (n + R E_0) - 2 \pi i \zeta_1 n}~,  \label{z11d}\ee
with
\be  R E_0 = {wR^2  \over \alpha'}(1 - b)\ .  \ee
A similar analysis for $w<0$ leads to the more general expression for $E_0$, which is valid for all $w\in \mathbb{Z}$,
\be R E_0 = {R^2  \over \alpha'}(|w| - bw)\ . \label{E0} \ee
For $b=0$, \eqref{E0} is identical to \eqref{zeromass}. And, as there, we find from \eqref{z11d} that 
\be R {\cal E} = R (E - E_0) =n~, \label{deltaE} \ee
in agreement with \eqref{eplus}. The $b$ dependence of $E_0$ \eqref{E0} was explained above.

To recapitulate, we find that the thermal partition sum of the $(0,2)$ string, \eqref{partsum}, reproduces the expected answer, 
\eqref{zthermal}, with the spectrum found in subsection \ref{spectrum}. Of course, this had to be the case, so the calculation of this section is a consistency check on the technology.

\section{Some properties of $\M^N/S_N$} \label{symmprodn}

In this section, we discuss the symmetric product of Niemeier CFT's, $\M^N/S_N$. This problem is of interest in its own right, but our motivation for studying it is that, as we will see in section \ref{spdyn}, it is directly relevant to the $(0,2)$ heterotic string studied in the previous sections. 

Symmetric product CFT's play an important role in matrix string theory \cite{Dijkgraaf:1997vv}, holography in $AdS_3$ and its single-trace $T\bar T$-type irrelevant deformations,\footnote{For a recent description of the status of symmetric product CFT's in $AdS_3$ and single-trace $T\bar T$, as well as references to earlier work, see \cite{Chakraborty:2023mzc}.} and have been extensively studied in these contexts. One new element here is that the seed CFT, $\M$, is holomorphic. We will discuss below some of the features particular to that case. 

\subsection{Spectrum}

The spectrum of $\M^N/S_N$ splits into different twist sectors. The untwisted sector consists of all operators $V$ in $\M$, summed over the different copies in the symmetric product. Thus, it is isomorphic to $\M$. In addition, there are $\mathbb{Z}_w$ twisted sectors, obtained by picking $w$ of the $N$ copies of the seed $\M$, and looking at states twisted by the $\mathbb{Z}_w$ that takes $\M_1\to\M_2\to \cdots \to\M_w\to \M_1$. As we will see below, the spectrum in these sectors takes the form 
\be 
\label{twistw}
\Delta_w={\Delta-1\over w}+w~,
\ee
where $\Delta$ runs over the spectrum of the seed CFT $\M$. For $w>1$, not all $\Delta$ contribute. The requirement that the dimensions $\Delta_w$ \eqref{twistw} are integer, which is necessary for consistency, implies that $\Delta-1\in w\mathbb{Z}$. 

In the symmetric product, twisted  sectors are labeled by conjugacy classes,
\be (w_1, w_2, w_3, \ldots w_m)~, \qquad \sum_{j=1}^m w_j = N \ ,  \label{sectors}\ee
which are obtained by dividing the $N$ copies of $\M$ into groups of $w_j$ copies, $j=1,\cdots, m$, and looking at 
$\mathbb{Z}_{w_j}$ twisted sectors for each group. We will see in section \ref{spdyn} that in the context of $(0,2)$ string theory, the $\mathbb{Z}_w$ twisted sector describes single-string states winding $w$ times around the spatial circle, while the general states \eqref{sectors} correspond to $m$ particle states, consisting of strings with windings $w_j$.

In the $\mathbb{Z}_w$ twisted sector, local operators $V$ in $\M$ transform non-trivially under a $2\pi$ rotation in the $x$-plane:
\be
\label{zwtrans}
V_I(e^{2\pi i}x)=V_{I+1}(x)~,
\ee
where $I=1,\cdots, w$ labels the $w$ copies of $\M$ that transform to each other under $\mathbb{Z}_w$ in the way indicated above, and $V_{w+1}(x)=V_1(x)$. One can think of \eqref{zwtrans} as due to an insertion of a twist field $\sigma_w$ at the origin of the $x$-plane. 

To compute the spectrum in the $\mathbb{Z}_w$ twisted sector, it is convenient to go to the covering space of the $x$-plane (see e.g. \cite{Roumpedakis:2018tdb}), and define a complex coordinate 
\be
\label{covspace}
t=x^{1\over w}~.
\ee
On the $t$ plane, the conditions \eqref{zwtrans} imply that the operators $V$ are single valued, and we have a single copy of the CFT $\M$. In this CFT, the operator $V$ creates a state with dimension $\Delta$. In order to compute the dimension of this state in the original CFT on the $x$-plane, we need to relate scaling dimensions on the covering space labeled by $t$ to those on the $x$-plane. Using the standard transformation of the stress-tensor under conformal transformations,
\be x'(t)^2T(x)=T^{\rm cov}(t)-{c\over12}\{x,t\}~,\;\;\;
\{x,t\}={x'''(t)\over x'(t)}-{3\over2}{x''(t)^2\over x'(t)^2}~,\ee 
one finds \cite{Roumpedakis:2018tdb}
\be\label{covsp}
L_0^{\rm cov}=wL_0-{c\over24}(w^2-1)~.
\ee
Plugging $c=24$, $L_0=\Delta_w$ and $L_0^{\rm cov}=\Delta$ into \eqref{covsp}, we find the relation \eqref{twistw}.

\subsection{Symmetries\label{sec52}}

In this subsection, we discuss the following question. Consider a symmetric product $\M^N/S_N$, where the seed CFT $\M$ has Kac-Moody and Virasoro symmetries with generators $K^a_n$ and $L_n$, respectively.\footnote{In most of this subsection we do not assume that $\M$ is holomorphic.} We would like to understand the fate of these symmetries in the symmetric product CFT. We will discuss this issue for the Kac-Moody case, and comment on the generalization to Virasoro. 

The CFT $\M^N$ has an affine Lie algebra  generated by 
\be
\label{KMsym}
K^a_n=\sum_{\alpha=1}^N K^a_{n,\alpha}~,
\ee
where $K^a_{n,\alpha}$ are the Kac Moody generators $K^a_n$ in the $\alpha$'th copy of the CFT $\M$. The generators $K^a_n$ \eqref{KMsym} have the following properties:
\begin{itemize}
\item They are invariant under $S_N$, and therefore survive the projection to $\M^N/S_N$.
\item They satisfy the same affine Lie algebra as in the seed CFT $\M$.
\item If the level of the affine Lie algebra in the seed CFT is $k$, the level of \eqref{KMsym} is $Nk$. 
\end{itemize}
Now, consider the action of the generators \eqref{KMsym} on the states described in the previous subsection. These states break up into sectors labeled by conjugacy classes \eqref{sectors}, and one can discuss the action of $K^a_n$ on a particular sector. 

For a specific choice of $w_j$ in \eqref{sectors}, one can write $K^a_n$ as a sum of contributions of the different groups of $w_j$ copies of $\M$, 
\be
\label{KMsymw}
K^a_n=\sum_{j=1}^m K^a_n(w_j)~,
\ee
where $K^a_n(w_j)$ is the contribution to the sum \eqref{KMsym} of the $w_j$ copies of $\M$ that participate in the cycle $\M_1\to\M_2\to \cdots \to\M_{w_j}\to \M_1$ discussed above. 

The level of the current algebra generated by $K^a_n(w_j)$ is $kw_j$. Of course, the total level is still $Nk$, due to \eqref{sectors}, but if we restrict attention to the action of $K^a_n$ on the cycle $w_j$, it acts as an affine Lie algebra of level $kw_j$. 

It is easy to generalize the above discussion to the Virasoro case. The total Virasoro generators are 
\be
\label{Lnsym}
L_n=\sum_{\alpha=1}^N L_{n,\alpha}~,
\ee
and the central charge of the Virasoro algebra \eqref{Lnsym} is $Nc$, $N$ times the central charge of the seed CFT $\M$. When acting on states in the sector \eqref{sectors}, one can write the $L_n$ \eqref{Lnsym} as a sum of contributions of the cycles $w_j$, with each cycle contributing central charge $cw_j$. 

We note for future reference, that if we restrict to the action of \eqref{Lnsym} on the cycle $w$, we can translate the equation for the scaling dimensions \eqref{twistw} to one for the corresponding energies on the cylinder, 
\be
\label{enww}
R{\cal E}=\Delta_w-{c_w\over 24}={\Delta-1\over w}~.
\ee
Of course, in \eqref{enww} we did restrict to the case of interest for us, and in particular set $c_w=24w$, $\bar c=0$.

The above discussion is reminiscent of what we found for $(0,2)$ strings in section \ref{comp}. As mentioned above, this is not an accident, since the free string Hilbert space has a symmetric product structure, with a cycle $w$ corresponding to a single-string state with winding $w$.

\subsection{Torus partition sum}

In this subsection, we discuss the torus partition sum of a CFT of the form $\M^N/S_N$, for a given seed CFT $\M$, not necessarily holomorphic.\footnote{Not to be confused with the worldsheet partition sum discussed in section \ref{torus}. In particular, the modulus $\tau$ that we will use in this section is the analog of $\zeta$ \eqref{zetaa} there, and not of $\tau$ \eqref{partsum}.} The  $\M^N/\mathbb{Z}_N$ orbifold was discussed in \cite{Klemm:1990df}. We will generally follow their notations, and will comment on the points where the treatment of $S_N$ and $\mathbb{Z}_N$ orbifolds differ (see  \cite{Dijkgraaf:1996xw,Hashimoto:2019hqo} for related discussions). After reviewing the construction for general seed CFT $\M$, we will apply the results to the case of a holomorphic $\M$, and see how \eqref{twistw}, \eqref{enww} can be read off from the partition sum.

We start with the case $N=2$, where $S_2 = \mathbb{Z}_2$. 
The partition sum of states in $\M^2=\M\times \M$ that are invariant under the $\mathbb{Z}_2$ that exchanges the two copies is 
\be 
\label{zinv}
Z_{\rm invariant}^{(2)}(\tau, \bar \tau) =  {1 \over 2}Z_\M(\tau, \bar \tau)^2  + {1 \over 2} Z_\M(2 \tau, 2 \bar \tau) \ . \ee
The first term on the r.h.s. of \eqref{zinv} accounts correctly for the contribution of states in $\M^2$, where we take distinct states from each copy, but undercounts states for which the contributions of the two copies are the same. The second term corrects for that. 

$Z^{(2)}_{\rm invariant}$ \eqref{zinv} is not modular invariant by itself, but it can be completed to a modular invariant function by defining
\be Z^{(2)}_{\rm new}(\tau, \bar \tau) = {1 \over 2}Z_\M(\tau, \bar \tau)^2 + {1 \over 2} Z_\M(2 \tau, 2 \bar \tau) + {1 \over 2}\sum_{j=0}^1  Z_\M \left({\tau+j \over 2}, {\bar \tau+j \over 2}\right) \ . \label{Z2new}\ee 
The expression (\ref{Z2new}) is the torus partition function of the ${\cal M}^2/S_2$ orbifold, 
\be Z_{{\cal M}^2/S_2}=Z_{\rm new}^{(2)} = Z^{(2)}_{\rm invariant}(\tau, \bar \tau) + Z^{(2)}_{\rm twist}(\tau, \bar \tau)~. \label{z2spec}\ee
It is identical to eq. (10) of  \cite{Klemm:1990df}, as well as eq. (25) of \cite{Hashimoto:2019hqo}, where $Z_{\rm new}^{(2)}(\tau, \bar \tau) = \Xi_2(\tau, \bar \tau)$. 
From \eqref{zinv} -- \eqref{z2spec}, we see that the quantity 
\be Z_{\rm twist}^{(w)} =  {1 \over w}\sum_{j=0}^{w-1}  Z_\M \left({\tau+j \over w}, {\bar \tau+j \over w}\right) \label{Ztwist} \ee
(for $w=2$; the general $w$ expression will be useful for $N>2$ below) is the contribution of the $\mathbb{Z}_2$ twisted sector of the orbifold. In the language of \eqref{sectors}, the two terms on the r.h.s. of \eqref{z2spec} correspond to the contributions of the conjugacy classes $(1,1)$ and $(2)$, respectively.

In the application to string theory (section \ref{spdyn}),  one can interpret the first term on the r.h.s. of eq. \eqref{z2spec} as the contribution of two string states, each wound once around a spatial circle, while the second term is the contribution of a single, doubly wound, string. The two-string contribution is symmetrized, since the strings are indistinguishable.

For $N=3$, one can proceed in a similar way. The contribution of $S_3$ invariant states in $\M^3$ is
\be 
\label{z3inv}
Z_{\rm invariant}^{(3)}(\tau, \bar \tau) = {1\over 6}Z_\M(\tau, \bar \tau)^3  + {1 \over 2} Z_\M(2 \tau, 2 \bar \tau) Z_\M(\tau, \bar \tau) + {1 \over 3} Z_\M(3 \tau, 3 \bar \tau)\ . \ee
The three terms on the r.h.s. have the same interpretation as before: the first term correctly accounts for the contributions of states in $\M^3$, where we take distinct states from all three copies, and the other two terms correct for the undercounting of states where either two of the three or all three are the same. 

Completing \eqref{z3inv} to a modular invariant function, one finds the partition sum of $\M^3/S_3$,\footnote{This equation is equivalent to eq. (26) in \cite{Hashimoto:2019hqo}. It does not appear in \cite{Klemm:1990df}, since that paper considered the $\mathbb{Z}_3$ orbifold.} 
\beq Z_{{\cal M}^3/S_3}(\tau, \bar \tau) &=& {1\over 6} Z_\M(\tau, \bar \tau)^3  + {1 \over 2} Z_\M(2 \tau, 2 \bar \tau) Z_\M(\tau, \bar \tau)+ {1 \over 3} Z_\M(3 \tau, 3 \bar \tau)\cr
&+&  {1 \over 2} \sum_{j=0}^1 Z_\M \left({\tau+j \over 2}, {\bar \tau+j \over 2}\right) Z_\M(\tau, \bar \tau) \cr 
&+& 
 {1 \over 3} \sum_{j=0}^2 Z_\M \left({\tau+j \over 3}, {\bar \tau+j \over 3}\right) \ .  \label{z3new}\eeq
The first line of \eqref{z3new} is the contribution of the untwisted sector of the orbifold (the conjugacy class $(1,1,1)$ in \eqref{sectors}), the second is the contribution of states in the sector $(2,1)$, that are products of $\mathbb{Z}_2$ twisted and untwisted states, and the third line corresponds to the sector $(3)$ of $\mathbb{Z}_3$ twisted states. In the application to string theory, they correspond to three singly wound strings, two strings with windings one and two, and a single string with winding three, respectively. 

For general $N$, one can extend the above construction using the technology of Hecke sums. As explained in \cite{Hashimoto:2019hqo}, one can define  quantities ${\cal Z}_N(\tau, \bar \tau)$, which are related  to $Z_{\M^N/S_N}(\tau, \bar \tau)$ (which were denoted by $\Xi_N(\tau, \bar \tau)$ in \cite{Hashimoto:2019hqo}), by
\be 1 + \sum_{N=1} \eta^N Z_{\M^N/S_N}(\tau, \bar \tau) = \exp \left( \sum_{N=1} \eta^N {\cal Z}_N(\tau, \bar \tau) \right) \ . \label{Xi} \ee 
Here $\eta$ is a formal expansion parameter. The quantities ${\cal Z}_N$ are related by the Hecke sum,
\be {\cal Z}_N(\tau, \bar \tau) = T_N \left[Z_{\M}(\tau, \bar \tau) \right]~, \label{hecke} \ee
to the partition function $Z_{\M}(\tau, \bar \tau)$ of the seed CFT. We will not review the definition of $T_N$ here; see equations (16) -- (22) in \cite{Hashimoto:2019hqo}. 

One can check that the quantities $Z_{\M^N/S_N}(\tau, \bar \tau)$ on the l.h.s. of \eqref{Xi} are indeed the partition sums of the orbifold $\M^N/S_N$, that they can be written as a sum over contributions of different sets of $w_j$ \eqref{sectors}, and that the contribution of the $\mathbb{Z}_w$ twisted sector to the partition sum is given by  $Z_{\rm twist}^{(w)}$  \eqref{Ztwist}.

So far, we discussed the case where the seed CFT $\M$ is arbitrary. We next turn to the case where it is a holomorphic CFT with $c=24$. The partition sum of $\M$, $Z_{\M}(\tau)$, is given in this case by eq. \eqref{ZNiemeier} (though, again, $\tau$ here is the modulus of the target-space torus, while there it was that of the worldsheet one). To calculate the partition sum of the $\mathbb{Z}_w$ twisted sector of $\M^N/S_N$, we substitute (\ref{ZNiemeier}) into (\ref{Ztwist}). The sum over $j$ constrains $\Delta-1$ to be an integer multiple of $w$,
\be \Delta-1 = nw~, \qquad n\in\mathbb{Z}_+~.
\label{constraint}\ee
Evaluating the sum in (\ref{Ztwist}), imposing the constraint \eqref{constraint}, one finds 
\be Z^{(w)}_{\rm twist}(\tau)
=\sum_{n\ge 0} D(1+nw) e^{-2 \pi \tau_2 n+ 2 \pi i \tau_1 n}~. \label{twistedN}\ee
Comparing \eqref{twistedN} to the general form of the thermal partition sum \eqref{zthermal}, we see that the spectrum of the twisted sector includes states with energy 
\be R{\cal E} = n = {\Delta-1 \over w}~, \ee
and degeneracy $D(\Delta)$, in agreement with what we found earlier \eqref{enww}. 

While not essential for the main line of development of this paper, we next briefly comment on the relation between the partition sum of $\M^N/S_N$ and the Klein $j$ function. Any unitary, modular invariant holomorphic CFT with central charge $c=24$, such as a Niemeier CFT $\M$, has a torus partition sum that is equal up to an additive integer to the Klein $j$ function (assuming a unique $SL(2,R)$ invariant vacuum, which is certainly the case for Niemeier CFT's), 
\be j(\tau)=q^{-1} + 744+196884 q+
21493760 q^2+864299970 q^3+20245856256 q^4 + \ldots~. \label{jtau}
\ee
More precisely, one has 
\be Z_{\M}(\tau) = j(\tau) + j_0~, \qquad j_0=24 (h+1)-744~,  \label{zj}\ee
where $h$ is the Coxeter number of the lattice \cite{Dixon:1988qd,Jankiewicz:2005rx}. 

The partition function of $\M^N/S_N$ must take the form 
\be Z_{\M^N/S_N}(\tau) = \sum_{p=0}^N a_p j^p, \ee
where $a_N=1$, and $a_p$ are integers that depend on $N$. We next present the values of these integers for $N=2,3$. 

For $N=2$, using (\ref{jtau}) and substituting (\ref{zj}) into (\ref{Z2new}),
leads to a Laurent expansion of $Z_{\M^2/S_2}$, which agrees, term by term, with the expression
\be Z_{\M^2/S_2}= j^2 + a_1 j + a_0 \ , \label{Z2newJ}\ee
where
\be 
\label{ztwocoeff}
a_1 = -744+j_0~, \qquad a_0 = 81000 + j_0(j_0+3)/2 \ . \ee
For $N=3$, we substitute (\ref{zj}) into (\ref{z3new}) and find that 
\be Z_{\M^3/S_3} = j^3  + a_2 j^2 + a_1 j + a_0~, \label{Z3newJ} \ee
where
\beq 
\label{zthreecoeff}
a_2 &=& -1488+j_0, \qquad a_1 =339582+{1 \over 2} j_0(j_0-1484), \cr
a_0 &=& 60774522 + {3 \over 2} j_0^2 + {1 \over 6} j_0 (j_0^2 + 3804992) \ . \eeq
Note that the constants $a_p$ \eqref{ztwocoeff}, \eqref{zthreecoeff} are integers, as expected. It should be possible to repeat this analysis for arbitrary $N$ by going back to the Hecke structure (\ref{hecke}), but we will not do it here.

\section{Spacetime theory}
\label{spdyn}

In sections \ref{noncomp} -- \ref{torus}, we described some properties of $(0,2)$ heterotic strings from the worldsheet point of view. The purpose of this section is to discuss the spacetime theory, based on what we found in those sections, and the results of section \ref{symmprodn}. 

Consider first the free theory (i.e. set the string coupling $g_s=0$). In section \ref{comp}, we found that the single-string spectrum of the spacetime theory is labeled by a winding number $w\in \mathbb{Z}$. States with $w>0$ are left-moving, while those with $w<0$ are right-moving. The sector with $w=1$ is described by a left-moving CFT $\M_L$, which is a spacetime version of the worldsheet CFT $\M$ that enters the construction of the model. For $w=-1$, we find a right-moving CFT, $\M_R$, which is a right-moving spacetime version of $\M$.  

The sectors with $|w|>1$ of the string theory exhibit a symmetric product structure. For positive $w$, we found that the single-string spectrum, \eqref{massplus}, \eqref{eplus}, \eqref{niem}, is the same as that of the $\mathbb{Z}_w$ twisted sector of the orbifold CFT $\left(\M_L\right)^w/\mathbb{Z}_w$, \eqref{twistw}, \eqref{enww}. Similarly, for negative $w$, the spectrum \eqref{massminus}, \eqref{eminus} agrees with that of the $\mathbb{Z}_{|w|}$ twisted sector of the orbifold CFT $\left(\M_R\right)^{|w|}/\mathbb{Z}_{|w|}$. 

Therefore, the symmetric product CFT 
\begin{equation}
\label{mln}
\left(\M_L\right)^{N_L}/S_{N_L}
\end{equation}
captures via \eqref{sectors} all the multi-string states with arbitrary positive windings $w_j$ and total winding $N_L$. In other words, it captures all the left-moving excitations on an arbitrary collection of wrapped fundamental $(0,2)$ strings with fixed total winding. Similarly, the symmetric product CFT $\left(\M_R\right)^{N_R}/S_{N_R}$ captures all the multi-string states with negative $w_j$ in \eqref{sectors}, or equivalently the right-moving excitations on an arbitrary collection of wrapped anti-strings with total winding $N_R$.  

The $w=0$ sector of the theory plays a special role in the construction. As we saw in section \ref{comp}, it contains left and right-moving copies of the chiral algebra of $\M$. This algebra includes  Kac-Moody, Virasoro, and higher spin operators, which we constructed and discussed in subsections \ref{KM}, \ref{Vir}, and \ref{HS}, respectively. These generators act on the states in the sectors with $w\not=0$, and organize them into representations. The left-moving symmetry generators act on the left-moving theory \eqref{mln}, while the right-moving generators act on its right-moving analog. The central extensions of the Kac-Moody and Virasoro algebras depend on the winding $w$, see \eqref{commutrel}, \eqref{plm}, \eqref{plw}, \eqref{formC}, \eqref{viralg}. We explained this fact from the point of view of the symmetric product theory \eqref{mln}. 

General states, that contain strings with total winding $N_L$ and anti-strings with total winding $N_R$, are described in the free theory by the CFT
\begin{equation}
\label{mllrr}
\left(\M_L\right)^{N_L}/S_{N_L}\times \left(\M_R\right)^{N_R}/S_{N_R}~.
\end{equation}
A natural question is whether there is a maximal value that $N_L$ and $N_R$ in \eqref{mllrr} can take. This is a non-perturbative (in the string coupling $g_s$) question. In the free theory, one can take $N_L$, $N_R$ to be arbitrarily large, but at finite $g_s$ one may expect an upper bound on them (a stringy exclusion principle). Since in this paper we discussed the perturbative theory, we will leave this issue for future work.   

Looking back at equation \eqref{massplus}, we see that the conformal field theory energy on the cylinder of the theory \eqref{mln} differs from that found in string theory by the additive factor, 
\begin{equation}
	\label{falpha}
	{R\over\alpha'}\sum_jw_j={N_LR\over\alpha'}~.
	\end{equation}
One can think of this as a vacuum energy density (or cosmological constant), given by $N_L/\alpha'$. In the spacetime field theory \eqref{mln} it is a non-universal contribution to the energy, but it is natural that the $(0,2)$ string theory, that provides a particular UV completion, fixes it (to the value \eqref{falpha}). 

So far we discussed the free string theory. The next question we need to address is that of string interactions. We will mostly leave this issue to future work, restricting here to some initial remarks. 

It is natural to expect that for states that include only strings (or only anti-strings), the description \eqref{mln} (or its right-moving analog) is exact. If both strings and anti-strings are present, as in \eqref{mllrr}, we expect interactions between the left and right-moving excitations living on them. To understand the kind of interactions that we expect, it is useful to look back at equation~\eqref{NGact}. There, we found that the non-compact theory includes what looks like a $T\bar T$ deformation that couples the left and right-moving components of the scalar fields $\phi^j$, with a $T\bar T$ coupling proportional to $g_s^2\alpha'$. 

Therefore, it is natural to expect  that string interactions give rise to a $T\bar T$ deformation of the above CFT, that acts non-trivially on states that contain both strings and anti-strings. In a sector with given $N_L$ and $N_R$, this interaction takes the form $tT\bar T$, where $t\sim \alpha' g_s^2$ and $T$,~$\bar T$ are the total stress-tensors of $\left(\M_L\right)^{N_L}/S_{N_L}$, $\left(\M_R\right)^{N_R}/S_{N_R}$, respectively. We next comment on some features of such interactions. 

In sectors of the Hilbert space that contain only left-moving, or only right-moving, excitations, the $T\bar T$ interaction does not have an effect. E.g. in the symmetric product CFT \eqref{mln}, with any $N_L$, the $T\bar T$ interaction vanishes, since $\bar T=0$. To see the effects of this interaction, we must take both $N_L$ and $N_R$ to be non-zero. 

Consider, for example, the case $N_L=N_R=1$, corresponding to two strings with $w=1$ and $w=-1$, respectively. In the free theory, this sector is described by the CFT 
\be
\label{mmllrr}
\M_L\times \M_R~.
\ee
The $T\bar T$ interaction has an effect on the spectrum and symmetries. The effect on the spectrum is familiar from \cite{Smirnov:2016lqw,Cavaglia:2016oda}, while that on the symmetries is the following. In the individual CFT's $\M_L$ and $\M_R$ \eqref{mmllrr}, we have Kac-Moody currents $K^a(x)$ and $\bar K^a(\bar x)$, that are holomorphic and anti-holomorphic, respectively. Expanding them in modes, we get the charges $K^a_n$ \eqref{contour} and $\bar K^a_m$ \eqref{contourplus}, that we found in section \ref{comp}. 

When we turn on the $T\bar T$ interaction that couples $\M_L$ and $\M_R$ in \eqref{mmllrr}, a straightforward calculation gives, to first order in the $T\bar T$ coupling $t$:
\be \partial_{\bar x} K^a(x)=t\partial_x\left(K^a\bar T(\bar x)\right)~.\ee
Thus, the current $K^a$, that is holomorphic in the undeformed theory, remains conserved, but ceases to be holomorphic, due to the $T\bar T$ coupling. It still gives a global conserved charge, but all the conserved charges \eqref{contour} with $n\not=0$ are broken by the interaction, since their conservation relies on the holomorphy of $K^a$. Similar comments apply to the right-moving conserved current $\bar K^a(\bar x)$, and its modes \eqref{contourplus}.   

The resulting picture is consistent with what we found in section \ref{comp}. In general correlation functions, that involve sectors of the Hilbert space with $N_L, N_R>0$, the affine Lie algebras are broken, while if either $N_L$ or $N_R$ vanishes, we can use the Kac-Moody (and higher spin) symmetry generators to study the (undeformed) amplitudes. Since two-point functions involve amplitudes with either $(N_L,N_R)=(2,0)$ or $(0,2)$, there is no violation of the affine Lie algebras in the free theory. Higher-point functions exhibit this breaking, but since they correspond to string interactions, one can think of them as $g_s$ effects.

\section{Discussion}\label{discuss}

In this paper, we studied the $(0,2)$ heterotic string, building on some work from the 1990's \cite{Ooguri:1991fp,Ooguri:1991ie,Kutasov:1996fp,Kutasov:1996zm,Kutasov:1996vh}. The main reason for our renewed interest in this model is that it has some features in common with certain holographic models in string theory. Therefore, studying heterotic $\N=2$  strings may be useful for improving our understanding of such models, and vice versa. One can also hope that a better understanding of these models will lead to progress in the program of~\cite{Kutasov:1996fp,Kutasov:1996zm,Kutasov:1996vh}, that suggested that related models may play a role in string duality. 

The main result of this paper is a description of the spectrum and symmetries of the single-string Hilbert space of a class of vacua of the $(0,2)$ string. Such vacua are labeled by a choice of a left-moving worldsheet CFT with $c=24$. An example of such CFT's is obtained by studying twenty four chiral scalars living on a Niemeier torus, and we restricted to this class. We further restricted to the $1+1$ dimensional vacua of the $(0,2)$ string. 

Compactifying the spatial direction in target space on a circle, we found an interesting spectrum. The Hilbert space splits into sectors with different values of the string winding $w\in\mathbb{Z}$. Sectors with $w>0$ contain left-moving excitations, while those with $w<0$ contain right-moving ones. These excitations fall into a symmetric product structure. For positive $w$, the single and multi-string  excitations are described by the symmetric product \eqref{mln}, where the seed of the symmetric product $\M_L$ is the Niemeier CFT that appears in the worldsheet construction. For negative $w$, one finds a right-moving analog of \eqref{mln}, with (the right-moving analog of) the same seed CFT. 

The sector with $w=0$ plays an interesting role in the theory. It contains symmetry generators, corresponding to the chiral algebra of the seed CFT in \eqref{mln} and its right-moving analog. In particular, we found left and right-moving Kac Moody \eqref{contour}, \eqref{contourplus} and Virasoro \eqref{Ln}, \eqref{newLn} algebras, as well as higher spin operators \eqref{higherS}, that are in one to one correspondence with local operators in the Niemeier CFT $\M$. The left (right) moving symmetry generators act on the sectors with $w>0$ ($w<0$) as spectrum generating algebras. 

These algebras have a number of interesting properties. One is that their central extensions are $w$-dependent when acting on the single-string states with $w\not=0$. Another is that while they act well on excitations of the same chirality, they do not act well on excitations of the opposite chirality. This is one of a number of indications that the full spacetime theory, that describes both left and right-moving excitations, is not a CFT. We proposed that it might be a $T\bar T$ deformation of such a CFT, with a coupling proportional to $g_s^2\alpha'$. 

One of the things that drew our attention to heterotic $\N=2$ strings is their similarities to string theory on $AdS_3$. Some examples of these similarities are the following. In our study of $(0,2)$ strings, we found that they have a symmetric product structure, \eqref{mln} for $w>0$, and its right-moving analog for $w<0$. In string theory on $AdS_3$, there is a similar structure associated with long fundamental strings \cite{Argurio:2000tb,Giveon:2005mi,Balthazar:2021xeh,Eberhardt:2021vsx}, which plays an important role in studying the dynamics. In both theories, $\mathbb{Z}_w$ twisted sectors of the orbifold correspond to single-string states with winding $w$, and general conjugacy classes \eqref{sectors} correspond to multi-string states with windings $w_j$. The seed of the symmetric product corresponds to the sector with $w=1$, and in both cases it can essentially be read off the worldsheet analysis~\cite{Seiberg:1999xz}.

Another similarity between $(0,2)$ string theory and $AdS_3$ is that both theories exhibit infinite-dimensional symmetry algebras with central charges that depend on the winding sector. The generators of these symmetries, which include Kac-Moody and Virasoro algebras, come from the zero-winding sector. Their construction in our model is similar to that in $AdS_3$. For example, eq. \eqref{contour} is similar to eq. (2.27) in \cite{Giveon:1998ns}, and the $w$-dependent central charge \eqref{plm} is similar to (2.31) in that paper. 

The fact that the central charges of Kac-Moody and Virasoro algebras depend on the sector was seen in section \ref{symmprodn} to have a natural interpretation in the symmetric product. The total central charge is always the same, due to the sum rule \eqref{sectors}, but when we restrict attention to single-string states, i.e. to a particular $w_j$ in \eqref{sectors}, we only see the contribution of that sector. This clarifies the interpretation of some results in \cite{Kutasov:1999xu,Giveon:2001up}, regarding the interpretation of the central terms in the various algebras that appear in string theory on $AdS_3$. 

Of course, the analogy between $(0,2)$ strings and string theory on $AdS_3$ is not perfect. In particular, while in the $(0,2)$ string all the excitations are chiral, in string theory on $AdS_3$ the generic states in the seed CFT are not. A related point is that while in the $(0,2)$ string one can think of the $w=0$ sector as containing only the chiral algebra of the spacetime theory, in $AdS_3$ this sector contains an infinite number of non-chiral operators.

In section \ref{spdyn}, we argued that string interactions may give rise to a $T\bar T$ deformation of the spacetime CFT that describes free $(0,2)$ strings. An analog of that in string theory on $AdS_3$ is the single-trace $T\bar T$ deformation, an irrelevant deformation of the boundary CFT that leads in the bulk to a deformed geometry that approaches near the boundary a linear dilaton background \cite{Giveon:2017nie}. The deformed theory retains a symmetric product structure (for long strings), but the seed of the symmetric product is ($T\bar T$) deformed. 

In $AdS_3$, the single-trace $T\bar T$ deformation changes the energies of the states from their original (CFT) values, but for (anti-) chiral states the change vanishes. It was argued in \cite{Georgescu:2022iyx,Du:2024bqk} that in these sectors one can define Kac-Moody and Virasoro algebras, that act well on chiral states, but not on anything else. This is similar to what we found for the $(0,2)$ string in section~\ref{comp}. Our analysis clarifies the role of these symmetries. They can be viewed as spectrum  generating algebras for the chiral sectors, which is very natural since these sectors don't feel the $T\bar T$ deformation. Therefore, one can organize their spectrum w.r.t. the undeformed Kac-Moody, Virasoro, etc in that sector. 

These algebras are also relevant for correlation functions of arbitrary operators in the left-moving theory \eqref{mln}, or its right-moving analog. General correlation functions, which involve both left and right-moving operators, are not usefully constrained by them.

Some of our results are also reminiscent of those in \cite{McGough:2016lol,Callebaut:2019omt}.
In subsection 2.6 of~\cite{McGough:2016lol}, it was shown that there is a relationship between the critical bosonic string on $\mathbb{R}_t\times \mathbb{S}^1\times M$, where $M$ is a generic CFT with $(c,\bar c)=(24,24)$, and a $T\bar T$ deformation of the CFT $M$, with coupling $\alpha'$. In a certain gauge, which restricts the theory to a sector with a singly wound string on the $\mathbb{S}^1$, the $T\bar T$ deformed CFT is described in terms of the worldsheet theory of this bosonic string. 
In \cite{Callebaut:2019omt}, it was shown that in the chiral sector of this theory, there is a conserved Virasoro symmetry (and similarly for the anti-chiral sector); see eq. (4.5) in \cite{Callebaut:2019omt}.\footnote{That paper considers a more general problem, where the string is non-critical. The analogy to our construction involves the  critical case.}  Comparing to our eq.~\eqref{Ln}, we see that the two constructions are similar. And, in both constructions, the (anti-) chiral Virasoro generators don't act well outside the sector of (anti-) chiral operators.

While in this paper we focused on perturbative string physics, we mentioned the idea that the full theory at a finite value of the string coupling $g_s$ may be a $T\bar T$ deformation of a CFT of the form 
\be 
\label{fullcft}
\left(\M_L\right)^N/S_N\times \left(\M_R\right)^N/S_N~,\ee
where $N\sim 1/g_s^2$ is the total number of (anti-) strings, and the $T\bar T$ coupling is $t\sim g_s^2\alpha'$. If the coupling $t$ is positive (in the usual conventions in the literature), the high-energy spectrum exhibits Hagedorn growth, with entropy (here and below we omit order one numerical constants)~\footnote{See e.g. \cite{Giveon:2017nie}.} 
\be S_H\sim\sqrt{\alpha'}E~.\ee
On the other hand, for energies that are low enough that we can ignore the $T\bar T$ interaction between left and right-movers, we find the usual Cardy entropy of the CFT \eqref{fullcft},
\be S_C\sim \sqrt{NRE}~.\ee
The crossover between the two behaviors happens at an energy $E_c$,
\be E_c\sim {NR\over \alpha'}~.\ee
This energy is comparable to the zero-point energy \eqref{falpha} that we found in the $(0,2)$ string analysis of section \ref{comp}. We view this agreement as a check of the basic idea that such a $T\bar T$ coupling between left and right-moving spacetime excitations is present in the $(0,2)$ string.

The results of this paper lead to many questions and avenues for further progress. We next list some of them:
\begin{itemize}
\item The worldsheet construction gives rise to a spacetime theory in which the left and right-moving wound strings are described by \eqref{mln} and its right-moving analog, with $\M_L$ and $\M_R$ being the same Niemeier CFT, $\M$. From the worldsheet point of view, the fact that $\M_L$ and $\M_R$ are the same is obvious -- the worldsheet construction takes as input one particular $\M$. However, from the spacetime point of view, it is not clear what would go wrong if, for example, we took $\M_L$ and $\M_R$ to be two different Niemeier CFT's. 
\item Another, perhaps related, consequence of the worldsheet construction, is the spacetime symmetry structure. We saw in subsection \ref{KM} that while the full affine Lie algebras, $\widehat G_L$ \eqref{contour} and $\widehat G_R$ \eqref{contourplus}, do not act well on the full Hilbert space of the theory, their zero modes $K^a_0$, $\bar K^a_0$ do. Moreover, they are equal to each other, $K^a_0=\bar K^a_0$. Thus, the spacetime theory has one global symmetry, $G$, as opposed to two -- one from the left-movers, and another from the right-movers. It would be interesting to understand this phenomenon from the spacetime point of view. 
\item We argued that the spacetime theory describing the left-moving states is the symmetric product CFT \eqref{mln}. In that theory, we can calculate the correlation functions of local operators. Presumably, if our expectation is correct, we should be able to compute these correlation functions directly in the $(0,2)$ string theory. It would be interesting to do it, especially given the fact that the standard string scattering amplitudes vanish in the $\N=2$ string.
\item The worldsheet construction of section \ref{comp} has a Narain moduli space, relating different even self-dual lattices of signature $(25,1)$ \cite{Polchinski:1998rq}. This moduli space is twenty five dimensional. Twenty four of the moduli are the objects $C^j_x$ in \eqref{defac1}, and the last one is the radius of the $x$ circle. As is familiar from other contexts, moving around in this moduli space interpolates between different Niemeier CFT's. It would be interesting to understand this interpolation from the spacetime point of view \eqref{mln}, \eqref{mmllrr}.
\item In this paper, we focused on the $1+1$ dimensional vacua of $\N=2$ heterotic strings. It would be interesting to generalize the discussion to the $2+1$ dimensional vacua. Since the $1+1$ dimensional vacua give rise to spacetime theories that are of more general interest in field and string theory, the $2+1$ dimensional vacua may also lead to interesting spacetime theories. Some comments on these models appear in the 1990's literature, but no systematic analysis seems to exist. We hope to perform such an analysis in future work. 
\item In this paper we studied $(0,2)$ strings. In \cite{Kutasov:1996fp,Kutasov:1996zm,Kutasov:1996vh} it was argued that the heterotic $(1,2)$ string leads to a particularly interesting class of theories. In particular, its $1+1$ dimensional vacua give rise to critical bosonic, type II and other strings in target space, whereas the $2+1$ dimensional vacua give rise to membrane worldvolume theories. This makes it interesting to generalize our discussion to that case. We hope to report on this generalization in future work. 
\end{itemize}

\noindent
{\bf Note added:} In a follow-up paper \cite{Giveon:2024vlh}, we calculated the correlation functions of left-moving operators in the $(0,2)$ string, and showed that they agree with the holomorphic CFT ones, as anticipated above.

\section*{Acknowledgements}

The work of AG was supported in part by the ISF (grant number 256/22).
The work of AH was supported in part by the U.S. Department of Energy, Office of Science, Office of High Energy Physics, under Award Number DE-SC0017647.
The work of DK was supported in part by DOE grant DE-SC0009924. We used AI (Claude) to do some of the calculations (in particular those in appendix B).

\appendix

\section{Virasoro Algebra} \label{apenvir}

The purpose of this appendix is to describe some properties of the operator $L_n$ \eqref{Ln}. The first is that this operator is primary under worldsheet Virasoro. As mentioned in section \ref{comp}, the first part of this operator is not primary. Indeed, using the OPE 
\be  x^+(z) x^-(\xi) \sim \alpha' \log (z-\xi) \ , \ee
and the form of the worldsheet stress-tensor \eqref{leftxmu}, $T_x={1\over\alpha'}\partial x^+\partial x^-$, we find that the OPE
\be T_x(z)\, {iR\over\alpha'}\partial_\xi x^+e^{i\frac nR x^-}(\xi,\bar\xi)={iR\over\alpha'^2} \partial_z x^+\partial_z x^- (z) \, \partial_\xi x^+e^{i\frac nR x^-} (\xi,\bar\xi)\ee
contains a cubic pole, with residue 
\be -ne^{i\frac nR x^-}(\xi,\bar\xi)~.\ee
The second contribution to \eqref{Ln} can be written as, \eqref{lndef},
\be {n\over2} {\partial^2 x^-\over\partial x^-}e^{i\frac nR x^-}(\xi,\bar\xi)~,\ee
and its OPE with $T_x(z)$ also has a cubic pole, with the opposite residue. Thus, in the full operator \eqref{Ln} the cubic pole cancels, and we conclude that this operator is primary under worldsheet Virasoro.

The second property of \eqref{Ln} that we would like to verify is the algebra 
\eqref{viralgggg}. If we again write $L_n$ as a sum of two terms,
\be L_n=L_n^{(1)}+L_n^{(2)}~,\ee
corresponding to the two terms on the r.h.s. of \eqref{Ln}, the commutator \eqref{viralgggg} contains four terms. We next discuss them in turn. 

The first term is itself a sum of two different terms, corresponding to different Wick contractions:
\beq 
\label{aone}
[L_n^{(1)}, L_m^{(1)}]_1 &=& \oint {d\xi \over 2 \pi i} \oint_\xi {dz \over 2 \pi i } \ 
{i R \over \alpha'}  \partial x^+ e^{i n x^-/R}(z) \ {i R \over \alpha'} \partial x^+ e^{i m x^-/R}(\xi) \cr
& = & {i R \over \alpha'}\oint {d\xi \over 2 \pi i} \oint_\xi {dz \over 2 \pi i } \ 
{i R \over \alpha'}{i m \alpha' \over R(z-\xi)} \partial x^+ e^{i(n+m) x^-/R}(\xi)\cr
& + & {i R \over \alpha'}\oint {d\xi \over 2 \pi i} \oint_\xi {dz \over 2 \pi i } \ 
{i R \over \alpha'}{i n \alpha' \over R(\xi-z)} \partial x^+ e^{i(n+m) x^-/R}(\xi) \cr
& = & {i R \over \alpha'}\oint {d\xi \over 2 \pi i} (n-m)  \partial x^+ e^{i(n+m) x^-/R}(\xi) \cr
& = & (n-m) L_{n+m}^{(1)}~,
\eeq
and
\beq 
\label{atwo}
[L_n^{(1)}, L_m^{(1)}]_2 &=& \oint {d\xi \over 2 \pi i} \oint_\xi {dz \over 2 \pi i } \ 
{i R \over \alpha'}{i m \alpha' \over R(z-\xi)} {i R \over \alpha'}{i n \alpha' \over R(\xi-z)}
 e^{in x^-/R}(z)  e^{im x^-/R}(\xi) \cr
&=& \oint {d\xi \over 2 \pi i} \oint_\xi {dz \over 2 \pi i } \ 
{i R \over \alpha'}{i m \alpha' \over R(z-\xi)} {i R \over \alpha'}{i n \alpha' \over R(\xi-z)}
 (z-\xi) {in \over R} \partial x^- e^{i(n+m) x^-/R}(\xi)\cr
&=& \oint {d\xi \over 2 \pi i} \ 
\left(-   {in^2 m \over R} \right)\partial x^- e^{i(n+m) x^-/R}(\xi)~.
\eeq
For $n+m \ne 0$, the operator on the r.h.s. of \eqref{atwo} vanishes, as discussed after eq.~\eqref{central}. For $n+m=0$, we find  
\be 
\label{athree}
\oint {d\xi \over 2 \pi i}  \ 
   {n^3 \over R} i \partial x^- (\xi) =n^3 P^-_L~,
\ee
where $P_L^-$ is the central operator \eqref{plm}.  Adding up the two contributions \eqref{aone}, \eqref{atwo}, we find that $L_n^{(1)}$ satisfies the algebra 
\begin{equation}
	\label{Lnmone}
	[L_n^{(1)},L_m^{(1)}]=(n-m)L_{n+m}^{(1)}+n^3P_L^-\delta_{n+m,0}~.
	\end{equation}
We next turn to the two cross-terms, 
$[L_n^{(1)},L_m^{(2)}]$, $[L_n^{(2)},L_m^{(1)}]$. For $n+m\not=0$, we have a calculation that is very similar to what we did above, so we will not review it.
It gives
\be [L_n^{(1)}, L_m^{(2)}]+[L_n^{(2)}, L_m^{(1)}]=(n-m)L_{n+m}^{(2)}~.\ee
For $n+m=0$, one has an additional contribution,
\beq 
[L_n^{(1)}, L_m^{(2)}] &=& \oint {d\xi \over 2 \pi i} \oint_\xi {dz \over 2 \pi i } \ 
{i R \over \alpha'} \partial x^+ e^{i n x^-/R} (z)\cdot{m \over 2} \cdot {\partial \partial x^- \over \partial x^-} e^{i m x^-/R} (\xi) \cr
& = &  \oint {d\xi \over 2 \pi i} \oint_\xi {dz \over 2 \pi i } \ {i R \over \alpha'} {2 \alpha' \over (z-\xi)^3} e^{i n x^-/R + i n (z-\xi)\partial x^-/R+ \ldots}(\xi) \ \cdot{m \over 2} \cdot {1 \over \partial x^-} e^{i m x^-/R}(\xi) \cr
& = &  \oint {d\xi \over 2 \pi i} \oint_\xi {dz \over 2 \pi i } \ {i R \over \alpha'} {2 \alpha'\over (z-\xi)^3} {1 \over 2} ( i n (z-\xi)\partial x^-/R)^2  \cdot{m \over 2} \cdot {1 \over \partial x^-} \cr
& = & \oint {d\xi \over 2 \pi i} \oint_\xi {dz \over 2 \pi i } \ {i  \over R} {1 \over (z-\xi)}   {n^3 \over 2} \partial x^-  \cr
& = &  \oint {d\xi \over 2 \pi i} \ {i n^3 \over 2 R}   \partial x^- ={n^3\over2} P^-_L~.
\eeq
Therefore, combining the two cross-terms, we find the following contribution to the $\delta_{n+m,0}$ term:
\be 
 [L_n^{(1)}, L_m^{(2)}]+  [L_n^{(2)}, L_m^{(1)}] = \delta_{n+m,0}\oint {d\xi \over 2 \pi i} \ {i n^3 \over R} \partial x^- =n^3 P_L^-~.\ee
This doubles the coefficient of $P_L^-$ in \eqref{athree}, \eqref{Lnmone}, and together with the other terms discussed above, gives \eqref{viralgggg}, \eqref{formC}.

\section{Relating \eqref{Ln} and \eqref{newLn}} \label{apenvirrr}

In this appendix we show that the two expressions for the spacetime Virasoro generators, \eqref{Ln} and \eqref{newLn}, differ by a BRST exact operator. The key object is the operator
\begin{equation}
f_n\equiv\frac{e^{i\frac nR x^-}}{\partial x^-}\ .
\label{eq:fn}
\end{equation}
Since $x^-$ has no singular OPE with itself, the only contractions of
$T_x=\frac1{\alpha'}\partial x^+\partial x^-$ with $f_n$ come from
$\partial x^+(z)$, which acts through
\begin{equation}
\partial x^+(z)\,\partial x^-(w)\sim\frac{\alpha'}{(z-w)^2}\ ,\qquad
\partial x^+(z)\,e^{i\frac nR x^-}(w)\sim\frac{in\alpha'}{R(z-w)}\,e^{i\frac nR x^-}(w)\ .
\label{eq:basic-contr}
\end{equation}
Thus, $f_n$ is a primary of $T^m=T_x+T_y$, \eqref{leftxmu}, \eqref{leftyi}, of
dimension $-1$. The conformal normal ordered product $(T_xf_n)(w)$, defined
as the $(z-w)^0$ term in the OPE $T_x(z)f_n(w)$, receives a contribution
from the free-field normal ordered product, $\frac1{\alpha'}\partial x^+
e^{i\frac nR x^-}$, and from re-expanding the uncontracted $\partial x^-(z)$
around $w$ in the singular terms:
\begin{equation}
(T_xf_n)=\frac1{\alpha'}\,\partial x^+e^{i\frac nR x^-}
-\frac12\,\frac{\partial^3x^-}{(\partial x^-)^2}\,e^{i\frac nR x^-}
+\frac{in}{R}\,\frac{\partial^2x^-}{\partial x^-}\,e^{i\frac nR x^-}\ .
\label{eq:Txfn}
\end{equation}
Since $T_y$ has no singular OPE with $f_n$, we also have
$(T_yf_n)=T_y\,e^{i\frac nR x^-}/\partial x^-$, which is the first term in
the integrand of (3.51).

We now use $(T_yf_n)=(T^mf_n)-(T_xf_n)$ in \eqref{newLn}. The two Schwarzian terms
in \eqref{newLn} combine with the second term in (\ref{eq:Txfn}) into
\begin{equation}
-\frac32\,\frac{\partial^3x^-}{(\partial x^-)^2}
+3\,\frac{(\partial^2x^-)^2}{(\partial x^-)^3}
=-\frac32\,\partial\!\left(\frac{\partial^2x^-}{(\partial x^-)^2}\right)\ .
\label{eq:schw-total}
\end{equation}
Integrating by parts against $e^{i\frac nR x^-}$, the total derivative in
(\ref{eq:schw-total}) gives $\frac{3n}2\,\partial\ln(\partial
x^-)\,e^{i\frac nR x^-}$, while the last term in (\ref{eq:Txfn}) gives
$-n\,\partial\ln(\partial x^-)\,e^{i\frac nR x^-}$. Altogether, we find
\begin{equation}
\begin{split}
L_n^{(3.51)}=&\oint\frac{dz}{2\pi i}\left(\frac{iR}{\alpha'}\,\partial x^+
+\frac n2\,\partial\ln(\partial x^-)\right)e^{i\frac nR x^-}
\;-\;iR\oint\frac{dz}{2\pi i}\,(T^mf_n)\\
&+\frac{3iR}{2}\oint\frac{dz}{2\pi i}\,\partial\!\left(\frac{\partial^2x^-}{(\partial x^-)^2}\,e^{i\frac nR x^-}\right)\ .
\end{split}
\label{eq:Ln-rewrite}
\end{equation}
The first term on the r.h.s.\ is \eqref{Ln}. The last term is the integral of a
total derivative of a single-valued operator, and vanishes.

It remains to show that the second term is BRST exact. For a primary $f$ of
dimension $h$, one has
\begin{equation}
\{Q,bf\}=(T^mf)+T^{gh}f-b\left(c\,\partial f+h\,\partial c\,f\right)\ .
\label{eq:Qbf}
\end{equation}
For $h=-1$, using $T^{gh}=-2\,b\partial c-\partial b\,c$, the ghost terms
combine into a total derivative,
\begin{equation}
\{Q,bf_n\}=(T^mf_n)-\partial\left(bc f_n\right)\ .
\label{eq:Qbfn}
\end{equation}
Therefore,
\begin{equation}
L_n^{(3.51)}=L_n^{(3.30)}-iR\,\Big\{Q,\oint\frac{dz}{2\pi i}\,b\,f_n\Big\}\ ,
\label{eq:Ln-equiv}
\end{equation}
so the two sets of generators are equivalent in BRST cohomology. The
relation (\ref{eq:Ln-equiv}) holds for all $n$, including $n=0$, so the two
expressions for $L_0$ agree before the shift by $C/24$ discussed around
\eqref{lzero}, \eqref{formL0}.

The relation (\ref{eq:Ln-equiv}) can be thought of as a consequence of the
Virasoro constraint $T_x+T_y\simeq0$: \eqref{newLn} is written in terms of the
stress tensor of the Niemeier CFT, $T_y$, while \eqref{Ln} is written in terms of
$\partial x^+$, using $T_x/\partial x^-=\partial x^+/\alpha'$ up to the
normal ordering terms in (\ref{eq:Txfn}). As in the main text, we have treated the operator $1/\partial x^-$ formally.
Near insertions of the operators (3.2) with $w>0$, $\partial x^-\sim
-iwR/z$, so $1/\partial x^-$ is regular and single-valued there.

\providecommand{\href}[2]{#2}\begingroup\raggedright\endgroup

\end{document}